\documentclass[final,3p,twocolumn,10pt,fleqn]{elsarticle}

\usepackage{amssymb}
\usepackage{amsmath}

\usepackage{hyperref} 
\usepackage{booktabs}
\usepackage{tabularx}
\usepackage{multirow}
\usepackage{float} 
\usepackage[ruled,linesnumbered]{algorithm2e}
\usepackage{graphicx}
\usepackage{subcaption}
\usepackage{wrapfig}
\usepackage{ragged2e}

\usepackage[singlelinecheck=false]{caption}

\journal{Future Generation Computer Systems}

\begin{document}

\begin{frontmatter}



\title{Carbon-aware decentralized dynamic task offloading in MIMO-MEC networks via multi-agent reinforcement learning}


\author[a]{Mubshra Zulfiqar} 
\author[b]{Muhammad Ayzed Mirza} 
\author[c]{Basit Qureshi}

\affiliation[a]{organization={School of Computer
Science and Artificial Intelligence},
            addressline={Wuhan University of Technology}, 
            city={Wuhan},
            postcode={430000}, 
            state={Hubei},
            country={China}}
\affiliation[b]{organization={School of Computer Science
and Information Engineering},
            addressline={Qilu Institute of Technology}, 
            city={Jinan},
            postcode={250200}, 
            state={Shandong},
            country={China}}
\affiliation[c]{organization={Department of Computer
Science},
            addressline={Prince Sultan University}, 
            city={Riyadh},
            postcode={12435}, 
            state={Riyadh Province},
            country={Saudi Arabia}}

\begin{abstract}

Massive internet of things microservices require integrating renewable energy harvesting into mobile edge computing (MEC) for sustainable eScience infrastructures. Spatiotemporal mismatches between stochastic task arrivals and intermittent green energy along with complex inter-user interference in multi-antenna (MIMO) uplinks complicate real-time resource management. Traditional centralized optimization and off-policy reinforcement learning struggle with scalability and signaling overhead in dense networks. This paper proposes CADDTO-PPO, a carbon-aware decentralized dynamic task offloading framework based on multi-agent proximal policy optimization. The multi-user MIMO-MEC system is modeled as a Decentralized Partially Observable Markov Decision Process  (DEC-POMDP) to jointly minimize carbon emissions and buffer latency and energy wastage. A scalable architecture utilizes decentralized execution with parameter sharing (DEPS), which enables autonomous IoT agents to make fine-grained power control and offloading decisions based solely on local observations. Additionally, a carbon-first reward structure adaptively prioritizes green time slots for data transmission to decouple system throughput from grid-dependent carbon footprints. Finally, experimental results demonstrate CADDTO-PPO outperforms deep deterministic policy gradient (DDPG) and lyapunov-based baselines. The framework achieves the lowest carbon intensity and maintains near-zero packet overflow rates under extreme traffic loads. Architectural profiling validates the framework to demonstrate a constant $O(1)$ inference complexity and theoretical lightweight feasibility for future generation sustainable IoT deployments.

\end{abstract}


\end{frontmatter}



\section{Introduction}
\label{introduction}

Future generation computer systems are increasingly characterized by latency-sensitive microservices and fully distributed edge computing infrastructures. Smart applications such as self-driving cars and facial recognition alongside smart energy systems and advanced healthcare create substantial processing needs for IoT devices with limited resources in edge environments \cite{hortelanoComprehensiveSurveyReinforcementlearningbased2023}. IoT applications requiring extensive computational resources escalate computing demands and render local execution progressively challenging \cite{chenDynamicTaskOffloading2023}. Projections indicate the number of connected devices will reach 32.1 billion by 2030 \cite{statista_iot_connections_2024}. While cloud-based architectures handle massive data effectively, the rapid growth of IoT and wireless traffic increases workload demands and causes higher latency along with energy consumption and carbon emissions. MEC addresses these challenges by enabling smart devices to offload tasks to nearby edge servers. This paradigm supports low-latency and energy-efficient processing in distributed infrastructures compared to traditional cloud-centric models \cite{huaEdgeComputingArtificial2023}.

However, determining optimal task-offloading strategies remains difficult due to dynamic network conditions and limited edge resources. DRL demonstrates potential for learning adaptive and real-time offloading policies \cite{ganesanDeepLearningReader2022,sattarNovelRMSDrivenDeep2025,tsietsoMultiInputDeepLearning2023}. Several studies have explored task offloading \cite{chenRealTimeOffloadingDependent2024,jiaoDeepReinforcementLearning2023,pandaEnergyEfficientComputationOffloading2023,renDeepReinforcementLearning2021,zhouDeepReinforcementLearning2022}. Unlike traditional cloud systems handling large monolithic jobs, future MEC environments process numerous small and sudden tasks termed atomic micro-services. Edge computing supports granular task offloading to reduce transmission latency and network congestion and response delay \cite{zhouReconstructedGraphNeural2024,liMultiagentReinforcementLearning2024}. Additionally, research suggests models for deciding how to offload tasks through FL with secure partial offloading and cloud-edge collaboration \cite{mukherjeeJointTimeEnergyefficient2026}. Such approaches leverage idle mobile resources via elastic time-slot mechanisms to mitigate spatiotemporal mismatches and lower energy consumption \cite{suCloudedgeCollaborativeTask2026}.

Despite these advancements, the increasing scale of smart device deployments introduces new challenges related not only to performance but also to environmental sustainability. The ICT sector accounts for 1.7\% of global greenhouse gas emissions and could triple by 2040 without mitigation \cite{ayers_ballan_gray_mcdonald_ict_emissions_2024}. In MU-MIMO systems, spatial multiplexing boosts throughput but increases inter-user interference (IUI), requiring higher transmission power and proportionally raising carbon emissions \cite{leeOptimizedPowerControl2023}. Existing work reduces power but overlooks this interference-carbon feedback loop. Although energy harvesting (EH) supports battery-limited IoT devices, prior studies focus on energy efficiency rather than carbon reduction \cite{vitelloMobilityDrivenEnergyEfficientDeployment2022,zhouECMSEdgeIntelligent2022,guoEnergyHarvestingComputation2022}. Sustainable edge intelligence must therefore move beyond energy saving toward carbon awareness, aligning computation with renewable availability. This introduces a complex trade off among MIMO interference management, carbon minimization, and EH-based device longevity that current heuristic and single-objective DRL methods fail to resolve.

Addressing this tripartite trade off among carbon intensity, MIMO interference, and battery lifespan is critical for both network longevity and environmental sustainability in next-generation systems. With the growth of decentralized edge nodes in 6G networks, the aggregate carbon footprint may offset ICT efficiency gains. This work directly supports the United Nations Sustainable Development Goal (SDG) 13 regarding climate action by suggesting a carbon-first offloading framework \cite{un_desa_goal13_2025}. Unlike traditional energy-efficient schemes that minimize battery usage alone, this approach explicitly minimizes the carbon cost of grid interaction. Such a strategy contributes to the eco-sustainable digital infrastructure required for next-generation smart cities and addresses the industrial and social implications of distributed computing expansion. Moreover, beyond simulation-based evaluation, the simulation-to-real gap is bridged through architectural complexity analysis and ONNX model profiling to validate practical feasibility.

\subsection{Motivation and contribution}

The primary focus of this research is to minimize carbon emissions and buffer delay and energy wastage under massive concurrent task arrivals. Most DRL-based offloading schemes concentrate on user-device power consumption and overlook both edge server computational energy and associated grid carbon emissions \cite{guoEnergyHarvestingComputation2022}. Consequently, these approaches fail to address system-level energy wastage and the carbon footprint of MEC platforms with EH technologies \cite{yangCarbonAwareDynamicTask2024}. Base station contributions to carbon emissions within MEC frameworks are frequently disregarded. Prior research adopts centralized global decision-making rather than decentralized site-level optimization which limits scalability for dense microservice environments \cite{birhanieOptimizedTaskOffloading2024}. Furthermore, most EH-MEC research has been limited to single-antenna systems. To the best of our knowledge, only a few works explore MIMO-EH integration despite the potential for MIMO to improve transmission efficiency through spatial diversity \cite{longPowerAllocationScheme2023,zhouDynamicComputationOffloading2021}.

This paper develops a decentralized MIMO-enabled MEC framework with hybrid energy harvesting to tackle combined carbon emissions and latency challenges. Unlike conventional DRL methods that suffer from state-space explosion, the proposed model addresses massive IoT connectivity using a decentralized execution paradigm. Each IoT device independently learns decentralized offloading and power allocation strategies using only local observations without prior knowledge of system parameters. The main contributions are summarized as follows:

\begin{enumerate}

    \item A MIMO-enabled MEC framework with hybrid energy harvesting tailored for massive IoT microservices is developed. Unlike complex monolithic workflows, the workload is modeled as complex workflow applications composed of high-throughput and atomic and delay-sensitive sensing streams. This architecture addresses the spatiotemporal mismatch between stochastic task arrivals and intermittent renewable availability in fully distributed edge environments.
    \item A novel optimization framework minimizes long-term average computation cost by jointly integrating carbon emission and task delay metrics. This study proposes CADDTO-PPO which is a multi-agent optimization framework utilizing DEPS. This approach allows a shared policy to be trained on aggregated agent experiences while enabling independent and real-time power control and offloading decisions at the device level. The framework exploits continuous action-space learning to minimize long-term computation costs without requiring global Channel State Information (CSI).
    \item The gap between theoretical DRL and practical deployment is bridged by validating framework feasibility through rigorous architectural profiling and ONNX model serialization. Extensive complexity analysis demonstrates that CADDTO-PPO incurs negligible inference overhead and constant $O(1)$ time complexity. This proves viability for resource-constrained hardware and real-time urgent computing in future generation decentralized deployments.
    
\end{enumerate}

The remainder of this paper is structured as follows. Section 2 outlines the current status of research. Section 3 describes the system model and formulates the joint task offloading and resource allocation problem. Section 4 explains the suggested decentralized DRL-based offloading framework. Section 5 evaluates performance through architectural profiling and comparative analysis with baseline algorithms. Section 6 concludes the work and highlights promising directions for future research.
\section{Related work}
\label{related_work}

\begin{table*}[!t]
\centering
\scriptsize 
\caption{Comparison of carbon-aware task offloading and resource management in MEC}
\label{tab:related_work_comparison}
\addtolength{\tabcolsep}{-4.2pt} 
\begin{tabularx}{\textwidth}{l@{\hskip 0.05in}cc c ccc c p{2.1cm} >{\RaggedRight\arraybackslash\hyphenpenalty=10000}X} 
\toprule
\multirow{2}{*}{Articles} & \multicolumn{2}{c}{Approach} & \multirow{2}{*}{\shortstack{MIMO\\MEC/EH}} & \multicolumn{3}{c}{Objectives (Min.)} & \multirow{2}{*}{\shortstack{Arch\\Profiling.}} & \multirow{2}{*}{Solution Type} & \multirow{2}{*}{Key Contributions \& Constraints} \\
\cmidrule(lr){2-3} \cmidrule(lr){5-7}
& Cent. & Decent. & & Carbon & Latency & Energy & & & \\
\midrule
This work & -- & $\checkmark$ & $\checkmark$ & $\checkmark$ & $\checkmark$ & $\checkmark$ & $\checkmark$ & CADDTO-PPO & Sustainable MIMO-MEC; Constraints: Renewable energy, MIMO interference \\ \addlinespace

\cite{mukherjeeJointTimeEnergyefficient2026} & -- & $\checkmark$ & $\times$ & $\times$ & $\checkmark$ & $\times$ & $\times$ & Fed. Learning & Secure partial offloading in FL; Constraints: Data privacy, security \\

\cite{suCloudedgeCollaborativeTask2026} & $\checkmark$ & -- & $\times$ & $\times$ & $\times$ & $\checkmark$ & $\times$ & MINLP/Greedy & Elastic time-slot allocation; Constraints: Device battery budget \\

\cite{maCarbonNeutralEdgeComputing2023} & $\checkmark$ & -- & $\times$ & $\checkmark$ & $\times$ & $\times$ & $\times$ & Lyapunov & Carbon-aware ML offloading; Constraints: Carbon emission budget \\

\cite{liuTaskGraphOffloading2024} & $\checkmark$ & -- & $\times$ & $\times$ & $\checkmark$ & $\times$ & $\times$ & DQN & Scheduling dependent task graphs; Constraints: Task dependencies \\

\cite{chenPriorityawareTaskOffloading2025} & $\checkmark$ & -- & $\times$ & $\times$ & $\times$ & $\times$ & $\times$ & M2AC & LEO satellite edge computing; Constraints: Task deadlines \\

\cite{liMultiagentDeepReinforcement2025} & -- & $\checkmark$ & $\times$ & $\checkmark$ & $\times$ & $\times$ & $\times$ & MAPPO & Multitask load balancing; Constraints: Queue stability \\

\cite{zhaoDynamicOffloadingResource2021} & -- & $\checkmark$ & $\times$ & $\times$ & $\checkmark$ & $\times$ & $\times$ & Lyapunov & Joint dynamic scheduling; Constraints: QoS demands \\

\cite{zhouDynamicComputationOffloading2021} & $\checkmark$ & -- & $\checkmark$ & $\times$ & $\checkmark$ & $\checkmark$ & $\times$ & Lyapunov & Dynamic-evolving MIMO-MEC; Constraints: CSI availability \\

\cite{jiangTaskOffloadingResource2024} & $\checkmark$ & $\checkmark$ & $\checkmark$ & $\checkmark$ & $\checkmark$ & $\checkmark$ & $\times$ & TOMAC-PPO & Joint decentralized scheduling; Constraints: Resource contention \\

\cite{chakrabortySustainableTaskOffloading2022} & $\checkmark$ & -- & $\times$ & $\times$ & $\checkmark$ & $\times$ & $\times$ & Genetic Alg. & GA-based offloading; Constraints: Sustainability ratio \\

\cite{yangJointAccessSelection2025} & -- & $\checkmark$ & $\times$ & $\times$ & $\checkmark$ & $\times$ & $\times$ & Dueling DQN & LEO satellite selection; Constraints: Communication constraints \\

\cite{chenDynamicTaskOffloading2023}  & $\checkmark$ & -- & $\times$ & $\times$ & $\times$ & $\checkmark$ & $\times$ & Stochastic Opt & Dynamic MEC task scheduling; Constraints: Battery and delay \\
\bottomrule
\end{tabularx}

\begin{flushleft}
\vspace{2pt}
\scriptsize
Note: Cent.: Centralized; Decent.: Decentralized; EH: Energy Harvesting; Arch Profiling: Architectural Profiling; Min.: Minimization.
\end{flushleft}
\end{table*}

Recent advancements in MEC have raised challenges in achieving low-carbon, sustainable systems through green energy integration. To address the spatiotemporal mismatch between renewable energy supply and the bursty nature of IoT microservices, current research can be divided into three areas: using carbon-aware computing with energy harvesting, MIMO-enabled edge systems, and decentralized learning for managing resources.

\subsection{Carbon-aware computing and energy harvesting in MEC}

A key challenge in future generation edge computing infrastructures is balancing harvested renewable energy with grid power to minimize carbon footprints. Early studies emphasized energy efficiency rather than carbon intensity \cite{renDeepReinforcementLearning2021,weiUserSchedulingResource2018}. For instance, \cite{sunEnergyEfficientTaskOffloading2021} optimized energy-efficient offloading but treated all energy sources equally, ignoring fluctuating grid carbon intensity. Subsequent works examined energy sharing and emission quantification \cite{yuLessCarbonFootprint2023,savazziEnergyCarbonFootprint2023}, yet largely assumed static environments and single-antenna systems, neglecting interference-carbon interactions. More recent efforts introduced elastic time-slot mechanisms \cite{suCloudedgeCollaborativeTask2026} and centralized queue management \cite{birhanieOptimizedTaskOffloading2024}, but scalability and infrastructure carbon impacts remain limited. Closest to this work, \cite{weiDeepReinforcementLearning2023} proposed a carbon-aware DDPG framework, yet it depends on non-renewable backup without dynamic carbon intensity adaptation. Critically, none integrate carbon-aware optimization with MIMO interference management and energy harvesting in decentralized settings. In contrast, this approach integrates carbon-aware optimization, MIMO interference management, and energy harvesting within a fully decentralized framework. By adopting a carbon-first strategy, high-carbon grid usage is penalized and harvested energy is prioritized for delay-sensitive atomic microservices, enabling scalable and sustainable edge intelligence.

\subsection{MIMO-enabled edge computing and interference management}

MIMO enhances throughput in MEC with spatial diversity but introduces severe IUI, complicating distributed resource management. Prior work modeled MIMO-MEC using Lyapunov optimization \cite{zhouDynamicComputationOffloading2021}, yet required centralized control and global CSI, leading to high signaling overhead in massive IoT. DRL-based approaches such as \cite{chenDecentralizedComputationOffloading2020} enable local execution but still depend on centralized training and gradient aggregation, incurring communication costs, privacy risks, and straggler effects. Although federated solutions improve scalability \cite{mukherjeeJointTimeEnergyefficient2026}, they overlook energy harvesting-interference coupling. Existing studies also assume grid-powered devices or ideal channels, neglecting fading, interference, and renewable variability \cite{guoEnergyHarvestingComputation2022}. Critically, increasing transmission power to counter IUI raises grid-dependent carbon emissions. In contrast, the proposed fully decentralized framework jointly optimizes MIMO power control, task offloading, and energy harvesting scheduling, satisfying SINR constraints while explicitly minimizing grid carbon intensity without global CSI exchange or base station aggregation.

\subsection{Decentralized DRL for task offloading}

DRL is widely adopted for dynamic task offloading due to its ability to handle high-dimensional environments \cite{zabihiReinforcementLearningMethods2024}. Value-based methods such as DQN have been applied to DAG scheduling \cite{liuTaskGraphOffloading2024}, but struggle with convergence in continuous action spaces needed for fine-grained power control. Multi-agent approaches like MAPPO improve partial offloading and queue stability \cite{liMultiagentReinforcementLearning2024}, yet primarily optimize energy and load balancing, overlooking dynamic grid carbon intensity. Other works consider hybrid renewable-grid supply under QoS constraints \cite{zhaoDynamicOffloadingResource2021}, but remain limited to single-antenna systems and ignore long-term renewable stability in MIMO environments with stochastic arrivals and interference. Moreover, most DRL-based studies rely solely on simulations and lack practical feasibility validation \cite{liuTaskGraphOffloading2024,liMultiagentReinforcementLearning2024}. To address these gaps, CADDTO-PPO with parameter sharing is proposed as a fully decentralized DRL framework for multi-user MIMO-MEC systems. It adopts a carbon-first reward to explicitly minimize emissions and validates practical feasibility through architectural profiling and ONNX model serialization. Table \ref{tab:related_work_comparison} summarizes comparisons with existing work across multiple dimensions.
\section{System model}
\label{system_model}

\begin{figure*}[!t] 
   \centering
    \includegraphics[width=0.9\linewidth]{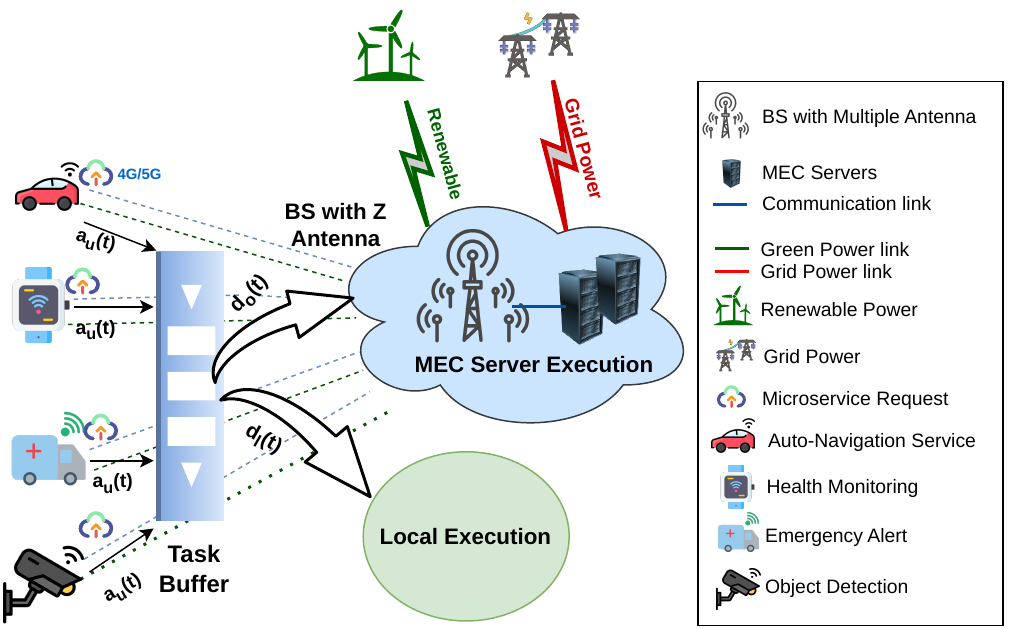}
    \caption{Framework of the carbon-aware MIMO-MEC system}
    \label{fig:system_model}
\end{figure*}

\begin{table}[!t]
\centering
\caption{Notations and descriptions}
\label{tab:notations}
\begin{tabularx}{\columnwidth}{l X} 
\toprule
\textbf{Notation} & \textbf{Description} \\
\midrule
$U$ & Set of IoT users \\
$T$ & Index set of time slots \\
$\delta$ & Length of one time slot \\
$c_u(t)$ & Processor computing speed for local execution during time slot $t$ of user $u$ \\
$N_u(C_u)$ & CPU cycles per task bit \\
$p_{l,u}(t)$ & Local power of user $u$ for local execution at slot $t$ \\
$d_{l,u}(t)$ & User $u$ processed data locally at slot $t$ \\
$p_{o,u}(t)$ & User $u$ transmission power for offloading at slot $t$ \\
$d_{o,u}(t)$ & Data offloaded by user $u$ at slot $t$ \\
$a_u(t)$ & Amount of task arrival (bits) of user $u$ at slot $t$ \\
$B_u(t)$ & Task queue buffer length of user $u$ at slot $t$ \\
$\Psi_u(t)$ & User $u$ receiving SINR to BS at slot $t$ \\
$r(t)$ & Received signal of BS at slot $t$ \\
$\beta_u$ & Normalized temporal channel correlation coefficient \\
$v_u(t)$ & Channel vector between user $u$ and BS at slot $t$ \\
$G_u^{(E)}(t)$ & Harvested energy of user $u$ at slot $t$ \\
$g_u(t)$ & Fraction of total energy consumption \\
\bottomrule
\end{tabularx}
\end{table}

In this section, the system model of a MIMO-enabled MEC network with EH is formulated.

\subsection{Network architecture}

An edge computing system is considered, comprising an MEC server co-located at a BS and multiple single-antenna IoT devices connected via uplink MIMO as illustrated in Fig. \ref{fig:system_model}. Each device harvests energy through an EH module via RF wireless transfer from an energy access point (EAP) at the BS, while the BS draws grid power to ensure task execution. A hybrid renewable-grid energy model enhances sustainability and extends device battery life. The BS, equipped with $Z$ antennas, supports high-density IoT connections by spatially multiplexing independent microservice streams. IoT users are denoted as $u \in U = \{1, 2, \dots, U\}$, and system operations follow a discrete-time model with equal-duration slots $\delta$ indexed by $\mathcal{T} = \{0,1,2,\dots\}$. Channel conditions and task arrivals vary over time ($t \in \mathcal{T}$), impacting transmission and offloading. Subsequent sections detail local/edge computing, communication, and computation models, with notations summarized in Table \ref{tab:notations}.

\subsection{Local computing}

The CPU of an IoT device serves as its core engine for local processing. For a given user $u \in U$, the processor's computing speed $c_u(t)$ (in cycles per second), which is controlled using dynamic voltage and frequency scaling (DVFS) techniques \cite{chenDecentralizedComputationOffloading2020} by tuning the chip voltage.

The computing speed is given by $c_u(t) = \sqrt[3]{\frac{p_{l,u}(t)}{k}}$, where $k$ denotes the effective switching capacitance of the processor. Here $p_{l,u}(t) \in [0, P_{l,u}]$ represents the local processing power allocated during time slot $t$, and $P_{l,u}$ being the maximum allowable local processing power. Accordingly, the CPU frequency is constrained as $c_u(t) \in [0, C_u]$, where $C_u = \sqrt[3]{\frac{P_{l,u}}{k}}$ represents the maximum permissible computing frequency for user $u$. Within time slot $t$, the locally processed bits by the IoT device are expressed as:
\begin{equation}
    d_{l,u}(t) = \delta \frac{c_u(t)}{N_u},
\end{equation}
where $N_u$ denotes the CPU cycles required per task bit, and its value could be determined by offline profiling or monitoring.

\subsection{Edge computing model}

In this model, computation tasks are offloaded from the IoT devices to the edge server. Let $p_{o,u}(t) \in [0, P_{u}^{\max}]$ denote the uplink transmit power of user $u$ at time slot $t$, where $P_{u}^{\max}$ represents the maximum allowable transmit power. Assuming the edge server possesses sufficient computational resources, the feedback delay is considered negligible for tasks generating small output data. 

The total volume of data (in bits) offloaded by user $u$ during time slot $t$ is determined by the achievable transmission rate $R_u(t)$ and the slot duration $\delta$:
\begin{equation}
d_{o,u}(t) = \delta \cdot R_u(t).
\end{equation}

The power consumption at the MEC server required to process these offloaded bits is modeled as a linear function of the data volume \cite{sunEnergyEfficientTaskOffloading2021}:
\begin{equation}
P_{\text{mec},u}(t) = \frac{\kappa N_u d_{o,u}(t)}{\delta},
\end{equation}
where $\kappa$ is the effective switched capacitance coefficient dependent on the MEC server's chip architecture, and $N_u$ denotes the number of CPU cycles required to process a single bit of data.

\begin{figure}[t]
   \centering
    \includegraphics[width=\linewidth]{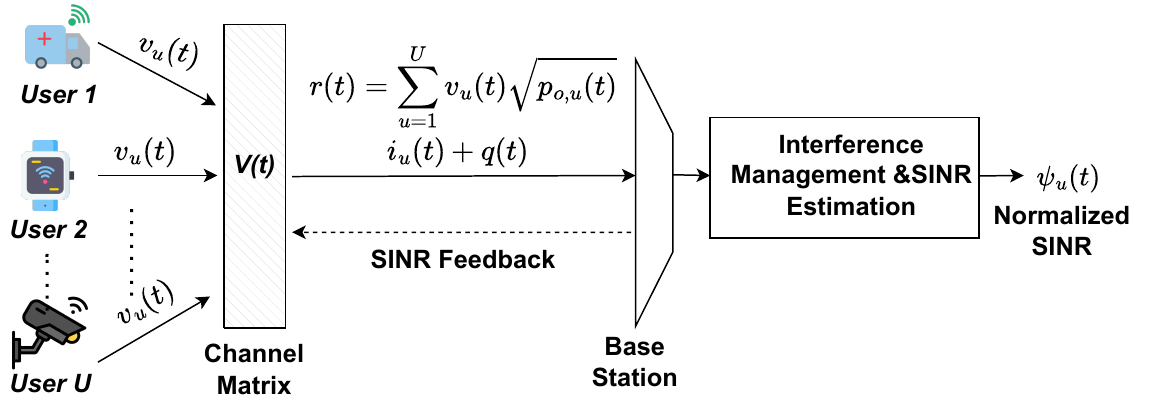}
    \caption{System architecture of the multi-user uplink MIMO-MEC system. The BS monitors SINR and provides feedback to decentralized agents for autonomous power control.}
    \label{fig:SINR}
\end{figure}

\subsection{Communication model}

For each user $u \in U$ during time slot $t$, the channel vector is $v_u(t) \in \mathbb{C}^{Z \times 1}$, representing a complex-valued vector of dimension $Z \times 1$ in the wireless communication system. Fig.\ref{fig:SINR} depicts the uplink of a multi-user MIMO system, where BS receives the signal $r(t)$. The received signal comprises the aggregate of transmitted data from all users, distorted by Additive White Gaussian Noise (AWGN). The uplink signal received at the BS is given by:
\begin{equation}
    r(t) = \sum_{u=1}^{U} v_u(t) \sqrt{p_{o,u}(t)} i_u(t) + q(t),
\end{equation}
where $i_u(t)$ denotes the complex data symbol transmitted by user $u$, assumed to have unit variance, and $q(t) \sim \mathcal{CN}(0, \sigma_R^2 I_Z)$ represents the AWGN vector, with zero mean and covariance $\sigma_R^2 I_Z$. Here, $I_Z$ is the $Z \times Z$ identity matrix. The noise vector $q(t)$ is assumed to be independent and identically distributed according to a complex Gaussian distribution.

The channel from user $u$ to the BS is modeled as a Gaussian-Markov block fading autoregressive process \cite{suraweeraEffectFeedbackDelay2011}. Over time, the channel vector $v_u(t)$ evolves as:
\begin{equation}
    v_u(t) = \beta_u v_u(t - 1) + \sqrt{1 - \beta_u^2} e(t),
\end{equation}
here, $e(t)$ is a complex Gaussian random vector, and $\beta_u$ is the temporal correlation coefficient of the user $u$ channel across consecutive time slots. This coefficient is expressed as $\beta_u = J_0(2\pi f_{d,u} \delta_t)$, based on Jake's fading spectrum \cite{suraweeraEffectFeedbackDelay2011}, where $f_{d,u}$ is the Doppler frequency of user $u$ and $J_0(\cdot)$ is the zeroth-order Bessel function of the first kind. 

Since all users transmit simultaneously over the same bandwidth $W$, the signal received at the BS from user $u$ is subject to co-channel interference from other active users. The SINR for user $u$ at time slot $t$ is defined as:

\begin{equation}
\psi_u(t) = \frac{p_{o,u}(t) |h_u(t)|^2}{\sigma_R^2 + \sum_{j \in \mathcal{U}, j \neq u} p_{o,j}(t) |h_{j,u}(t)|^2},
\end{equation}

\noindent where $|h_u(t)|^2$ denotes the channel gain of the intended user $u$, and $|h_{j,u}(t)|^2$ represents the interference channel gain from user $j$ to the BS. Consequently, the achievable data rate is given by Shannon capacity formula:

\begin{equation}
R_u(t) = W \log_2\left(1 + \psi_u(t)\right).
\end{equation}

\subsection{Computation model}

Unlike prior DAG-focused work suited for monolithic cloud applications, this study targets massive IoT microservices. In MEC scenarios like smart city surveillance and autonomous vehicle telemetry, workloads consist of bursty, independent, delay-sensitive tasks rather than sequential workflows.In such high-throughput environments, the main bottlenecks are wireless interference (MIMO constraints) and limited edge energy, not task dependencies. Task arrivals are modeled as a stochastic Poisson process to evaluate system resilience under high concurrency. Let $a_u(t)$ denote the number of task bits arriving for user $u$ in slot $t$, which are queued and executed in the subsequent slot $t + 1$. These arrivals are assumed to be independent and identically distributed (i.i.d.) across slots and follow a Poisson distribution, i.e., $a_u(t) \sim \text{Pois}(\lambda_u)$, where the mean $\lambda_u = \mathbb{E}[a_u(t)]$ represents the task arrival rate.

$B_u(t)$ represent the task queue buffer maximum size for user $u$ in time slot $t$. The number of task bits executed locally and offloaded is denoted $d_{l,u}(t)$ and $d_{o,u}(t)$, respectively. The overall bits processed during slot $t$ are $D_u(t) = (d_{l,u} + d_{o,u}) \leq B_u(t)$, where $B_u(t) \in [0, B_{max}(t)]$ and $B_{max}(t)$ represents the maximum buffer size. The queue evolution of user $u$'s task buffer over time is given by:
\begin{equation}
    RB_u(t) = \left[ B_u(t) - (d_{l,u}(t) + d_{o,u}(t)) \right]^+.
\end{equation}
\begin{equation}
    B_u(t + 1) = RB_u(t) + a_u(t), \quad \forall t \in \Delta,
\end{equation}

$RB_u(t)$ indicates the remaining buffer level, and the operator $[m]^+ = \max(m, 0)$ ensures non-negativity. This model accounts for offloading delays by incorporating time-dependency in buffer dynamics, adjusting the buffer length at $t + 1$ based on slot $t$ operations. This enables the prediction of variable processing latency, which depends on the volume of data under evaluation. The initial buffer condition is given by: $B_u(0) = 0$.

To monitor buffer overflows (i.e., when incoming data exceeds the buffer's storage or processing capacity), the buffer overflow metric is defined as:
\begin{equation}
    BO_u(t) = \max(0, RB_u(t) - B_{max}(t)).
\end{equation}

\subsection{Carbon emission model}

Let $G_u^{(E)}(t)$ denote the amount of green (harvested) energy available to user $u$ in time slot $t$. It is formulated as a Poisson-distributed random variable, $G_u^{(E)}(t) \sim \text{Pois}(\lambda_u^{(E)})$, where the mean value $\lambda_u^{(E)} = E[G_u^{(E)}(t)]$ denotes the expected quantity of harvested energy. For a more realistic energy model, both the BS and IoT devices in the network are assumed to utilize grid energy in addition to harvested green energy. When the harvested energy is insufficient, the grid energy acts as a backup supply. The total energy consumed by user $u$ at time slot $t$ is calculated as:
\begin{equation}
    E_u(t) = p_{l,u}(t) + p_{o,u}(t).
\end{equation}

The user prioritizes the use of harvested energy stored in a battery. Let $g_u(t) \in [0, 1]$ denotes the fraction of total energy consumption powered by renewable (green) energy for user $u$ in slot $t$. This fraction is determined by the available battery level $E_{battery,u}(t)$ relative to demand:
\begin{equation}
    g_u(t) = \min \left( 1, \frac{E_{battery,u}(t)}{E_u(t)} \right).
\end{equation}

Consequently, the battery state evolves based on the harvested energy and the portion of energy drawn from the battery $E_u(t) * g_u(t)$. The energy level of the user $u$ EH module battery at time $t$ is bounded by the battery capacity $B_{max}$:
\begin{equation}
\begin{aligned}
    E_{battery,u}(t+1) = \min [& E_{battery,u}(t) - E_u(t)g_u(t) \\
    & + G_u^{(E)}(t), B_{max}].
\end{aligned}
\end{equation}

When $g_u(t) < 1$, the remaining energy demand is drawn from the power grid. Additionally, the corresponding computation power consumed at the MEC server for the processing of user $u$ task, denoted as $P_{mec,u}(t)$, is entirely drawn from the grid. The total grid power consumption of the user $u$ during time slot $t$ is:
\begin{equation}
    E_{grid,u}(t) = E_u(t)(1 - g_u(t)) + P_{mec,u}(t).
\end{equation}

To support the decentralized carbon-aware task offloading framework, the individual carbon emission contribution of user $u$ at time slot $t$ is defined, denoted as $\varsigma_u(t)$. This value serves as the input for the agent's reward function:
\begin{equation}
    \varsigma_u(t) = E_{grid,u}(t) * e,
\end{equation}

Where $e$ denote the carbon emission factor. While real-world grid intensity varies, we use a mean carbon intensity $e = 700$ $\text{gCO}_2/\text{kWh}$, representing high-intensity fossil generation (e.g., natural gas or coal) \cite{IPCC2022AnnexIII}, to isolate the energy-latency trade-off and provide a worst-case baseline for carbon-aware decisions. For global evaluation and SDG 13 benchmarking \cite{UNStats2026}, the total system carbon at slot $t$ is the sum of all individual user contributions:
\begin{equation}
    C(t) = \sum_{u=1}^{U} \varsigma_u(t).
\end{equation}

\subsection{Energy wastage model}

To quantify inefficient energy usage, the Energy Wastage $EW_u(t)$ is defined. This metric quantifies the energy lost due to over-provisioning, where the agent allocates power for the data capacity $D_u(t)$ that exceeds the available buffer demand $B_u(t)$. It is calculated as:
\begin{equation}
    EW_u(t) = E_u(t) \times \frac{\max(0, D_u(t) - B_u(t))}{D_u(t)},
\end{equation}
Here, the term $\max(0, D_u(t) - B_u(t))$ represents the empty cycles or unused bandwidth.

\subsection{Problem formulation}

Based on the above discussion, the goal is to minimize total system carbon emissions and optimize both buffer delay and energy wastage for each IoT user. Consequently, the long term average computing overhead for user $u$ is defined as follows:
\begin{equation}
    G_u = \frac{1}{T} \sum_{t=0}^{T} w_{u,1} B_u(t) + w_{u,2} \varsigma_u(t) + w_{u,3} EW_u(t),
\end{equation}
where $w_{u,1}, w_{u,2}, w_{u,3} \in [0, 1]$ are weighted coefficients reflecting the trade-offs among carbon emission and delay and system inefficiency. The system seeks to minimize the computing overhead:
\begin{subequations} \label{eq:total_formulation}
\begin{align}
    & \min_{\{ p_{l,u}(t), p_{o,u}(t), g_u(t) \}} G_u \tag{\theparentequation} \label{eq:objective} \\
    \text{s.t.} \quad & C_1: p_{l,u}(t) \in [0, P_{l,u}], \quad \forall t \in \Delta, \label{eq:C1} \\
    & C_2: p_{o,u}(t) \in [0, P_{o,u}], \quad \forall t \in \Delta, \\
    & C_3: 0 \leq g_u(t) \leq 1, \\
    & C_4: E_u(t) \cdot g_u(t) \leq E_{battery,u}(t), \\
    & C_5: 0 \leq f_{mec} \leq f_{max},
\end{align}
\end{subequations}

Constraint $C_1$ limits each user local computing power $p_{l,u}(t)$ by $P_{l,u}$. $C_2$ bounds IoT transmission power. $C_3$ restricts the fraction of energy drawn from the grid. $C_4$ ensures renewable energy usage by user $u$ at timeslot $t$ does not exceed stored battery energy. $C_5$ caps the MEC server computation frequency. The optimization problem \eqref{eq:total_formulation} couples transmission power and offloading decisions under stochastic arrivals and time-varying channels to form an NP-hard MINLP problem. Traditional methods require global channel state information for MIMO interference management which incurs prohibitive overhead in massive IoT networks. To address this, a decentralized DRL approach is adopted. Each user $u$ makes decisions based solely on local observations. At the beginning of each time slot $t$, user $u$ observes incoming bits $a_u(t)$ and learns prior SINR $\Psi_u(t - 1)$ to estimate its current channel vector $v_u(t)$ using channel reciprocity. Using this state, the user independently determines local and offloading power without global coordination to avoid the latency and signaling overhead of centralized schemes.
\section{Carbon-aware decentralized dynamic task offloading algorithm}
\label{algorithm}

A carbon-aware dynamic decentralized task offloading (CADDTO) algorithm based on multi-agent proximal policy optimization (MAPPO) is proposed for MIMO-enabled MEC systems with hybrid energy supply. The framework jointly minimizes long-term computing cost and carbon emissions and buffer delay by penalizing energy wastage $EW_u(t)$ and buffer overflow $BO_u(t)$. A composite objective function is constructed using a weighted-sum approach, enabling flexible trade-offs between emission reduction and delay control \cite{songCarbonAwareFrameworkEnergyEfficient2024}. The state and action and reward spaces are designed to support decentralized decision-making.

Unlike centralized DRL schemes, a multi-agent PPO framework with parameter sharing is adopted. A single shared policy network $\pi_{\phi}$ is trained using aggregated trajectories while each user $u$ independently maps its local observation $s_u(t)$ to actions $a_u(t)$ during execution. This centralized training decentralized execution (CTDE) design ensures scalability in massive IoT networks and maintains constant model complexity and eliminates global information exchange during inference to significantly reduce signaling overhead.

\begin{figure}[t]
    \centering
    \includegraphics[width=\linewidth]{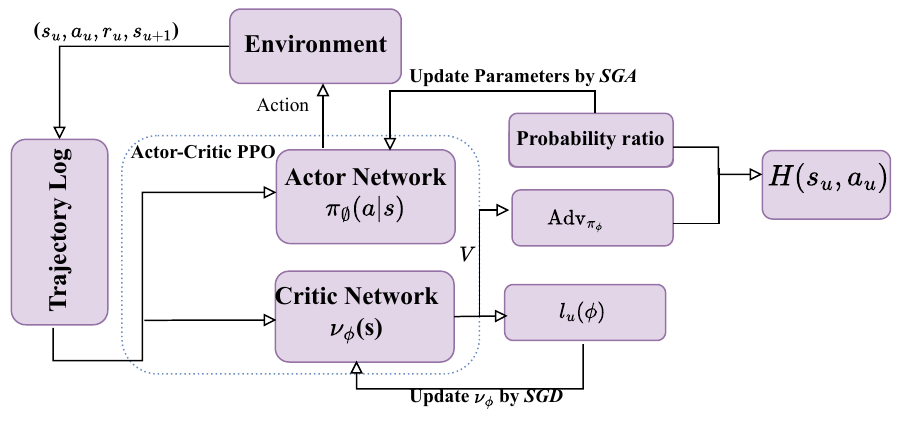}
    \caption{Multi-agent PPO architecture with centralized training decentralized execution using shared parameters}
    \label{fig:PPO}
\end{figure}

The PPO algorithm is an on-policy actor-critic method chosen for two reasons in green MIMO-MEC systems \cite{schulmanProximalPolicyOptimization2017}. First, unlike value-based methods that require discretized power levels and incur quantization errors, PPO handles continuous actions to enable fine-grained power control for carbon emission reduction. Second, unlike off-policy methods such as DDPG or SAC that rely on replay buffers, PPO uses fresh trajectories to ensure robust convergence in decentralized and non-stationary environments with rapidly changing interference. Fig. \ref{fig:PPO} depicts the actor-critic network of PPO.

To ensure scalability for massive IoT networks, parameter sharing is adopted by training a single actor network $\theta$ and critic network $\phi$ shared by all agents. Experiences from all users update these networks during training while execution remains fully decentralized. The offloading policy $\pi_\theta$ is optimized by maximizing the PPO objective function:

\begin{equation}
\theta^* = \arg \max_{\theta} \mathbb{E} \left[ L^{CLIP}(\theta) - c_1 L^{VF}(\theta) + c_2 S[\pi_\theta] \right],
\end{equation}
where $L^{CLIP}$ is the surrogate policy objective, $L^{VF}$ is the value function loss, and $S$ denotes the entropy bonus to encourage exploration, weighted by coefficients $c_1$ and $c_2$.

To reduce the variance of gradient estimates in the highly stochastic wireless environment, we employ Generalized Advantage Estimation (GAE). The advantage $\hat{A}_u(t)$ for user $u$ at time $t$ is calculated as:
\begin{equation}
\hat{A}_u(t) = \sum_{l=0}^{T-t-1} (\gamma \lambda)^l \delta_u(t+l),
\end{equation}
where $\gamma$ is the discount factor, $\lambda$ is the GAE smoothing parameter, and $\delta_u(t) = r_u(t) + \gamma V_\phi(s_u(t+1)) - V_\phi(s_u(t))$ is the Temporal Difference (TD) error.

The PPO surrogate objective with clipping is defined as:
\begin{equation}
\label{eq:ppo_clipped}
\begin{split}
    L^{CLIP}(\theta) = \mathbb{E} \Big[ \min \Big( & r_t(\theta) \hat{A}_t, \\
    & \text{clip}(r_t(\theta), 1-\epsilon, 1+\epsilon) \hat{A}_t \Big) \Big],
\end{split}
\end{equation}
where $r_t(\theta) = \frac{\pi_\theta(a_u|s_u)}{\pi_{\theta_{old}}(a_u|s_u)}$ is the probability ratio. The clipping parameter $\epsilon$ (set to 0.2) prevents destructive policy updates. 

The critic network estimates the long-term expected reward and is updated by minimizing the Mean Squared Error (MSE) loss:
\begin{equation}
L^{VF}(\phi) = \mathbb{E} \left[ (V_\phi(s_u) - R_u^{\text{target}})^2 \right],
\end{equation}
where $R_u^{\text{target}}$ is the computed return. The shared parameters are updated using Stochastic Gradient Descent (SGD) with the Adam optimizer, utilizing mini-batches sampled from the aggregated trajectories of all users. Algorithm 1 provides the corresponding pseudocode

\begin{algorithm}[tb]
\footnotesize
\caption{Decentralized CADDTO-PPO with Parameter Sharing}
\label{alg:CADDTO}
\KwIn{$N_{\text{user}}$, Max Steps $T_{\max}$, Episodes $Ep_{\max}$, Clip $\epsilon$, Learning Rates $\alpha, \beta$}
\textbf{Initialization:} Initialize shared Actor $\pi_{\phi}$ and Critic $V_{m}$ networks with random weights\;

\For{$\text{episode} = 1$ \KwTo $Ep_{\max}$}{
    $s \leftarrow \text{env.reset()}$\;
    \For{$t = 1$ \KwTo $T_{\max}$}{
        \For{each agent $u \in \{1, \dots, N_{\text{user}}\}$ \textbf{in parallel}}{
            \tcp{Decentralized Execution}
            Observe local state $s_u(t)$\;
            Sample action $a_u(t) \sim \pi_{\phi}(\cdot | s_u(t))$ (Stochastic)\;
        }
        Execute joint actions $\mathbf{a}(t)$ in Environment\;
        Observe rewards $r_u(t)$ and next states $s_u(t+1)$\;
        Store transitions $(s_u, a_u, r_u, s'_u, \log \pi(a_u))$ in shared buffer\;
    }
    
    \tcp{Centralized Training / Parameter Update}
    Compute Generalized Advantage Estimation (GAE) $\hat{A}$\;
    \For{$\text{epoch} = 1$ \KwTo $K$}{
        Sample mini-batches from shared buffer\;
        Compute ratio $\rho_t(\phi) = \frac{\pi_{\phi}(a|s)}{\pi_{\phi_{\text{old}}}(a|s)}$\;
        Compute Surrogate Loss: \\
        $L^{\text{CLIP}} = \min\left(\rho_t \hat{A}, \text{clip}(\rho_t, 1-\epsilon, 1+\epsilon)\hat{A}\right)$\;
        Update $\phi$ via Adam Optimizer\;
        Update Critic $m$ by minimizing $(V_m(s) - R_{\text{target}})^2$\;
    }
}
\end{algorithm}

\begin{figure}[t]
    \centering
    \includegraphics[width=\linewidth]{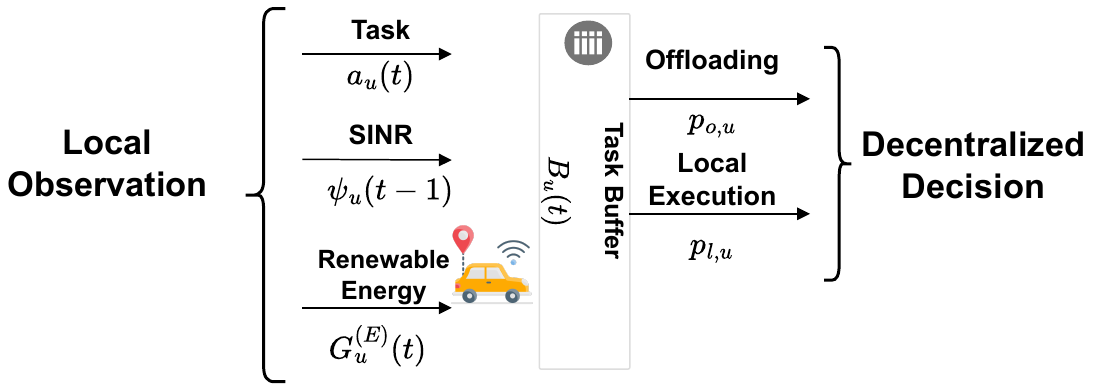}
    \caption{Schematic representation of the local observation and decentralized decision-making process of user agent at slot $t$}
    \label{fig:local}
\end{figure}

\textbf{State Space:} The state space of each user in the decentralized policy framework relies solely on local observations, as illustrated in Fig. \ref{fig:local}, it includes the previous time slot's normalized SINR, denoted as $\varphi_u(t-1)$, which is estimated from the received SINR $\psi_u(t-1)$, the current buffer queue length $B_u(t)$, and the quantity of harvested renewable energy $G_u^{(E)}(t)$. At the beginning of time slot $t$, the buffer queue length is refreshed based on the queue dynamics. 

The state space is designed to be minimal and locally observable to facilitate decentralized decision-making. To ensure numerical stability and faster convergence of the PPO algorithm, state variables are normalized to the range $[0,1]$. The local state $s_u(t)$ for user $u$ consists of:
\begin{itemize}
    \item Normalized Buffer Level: $\bar{B}_u(t) = B_u(t) / B_{\max}$, indicating the urgency of the computing task.
    \item Normalized SINR: $\bar{\psi}_u(t-1) = \text{clip}(\psi_u(t-1) / \psi_{\text{target}}, 0, 1)$, which provides an estimate of the local interference environment based on the feedback from the previous slot.
\end{itemize}
Unlike approaches requiring global CSI or zero factor precoding matrices, this state space relies solely on scalar feedback, significantly reducing signaling overhead. The state vector is defined as:
\begin{equation}
s_{u,t} = \{ \bar{B}_u(t), G_u^{(E)}(t), \bar{\psi}_u(t-1) \}.
\end{equation}

\textbf{Action space:} According to the current state of space $s_{u,t}$ and locally observed environment, for each user agent $u$, the action space is expressed as:
\begin{equation}
    a_{u,t} = [p_{l,u}(t), p_{o,u}(t)].
\end{equation}

Where $a_{u,t}$ represents the power allocation decision at time slot $t$ for both local execution and computation offloading. Specifically, $p_{l,u}(t) \in [0, P_{l,u}]$ and $p_{o,u}(t) \in [0, P_{o,u}]$. This action space offers a direct solution for decentralized dynamic computation offloading problems in the IoT environment, enabling each user to autonomously determine an optimal power allocation strategy.

\textbf{Reward:} The multi-objective reward function in the carbon-aware dynamic decentralized task offloading framework for MIMO-enabled MEC aims to optimize the long-term average computing overhead $G_u$ for each user $u$. Accordingly, the immediate reward $r_u(t)$ received by user $u$ in time step $t$ is given by:

\begin{equation}
\label{eq:reward_function}
\begin{split}
    r_u(t) = - \Big( & w_1 \frac{B_u(t)}{B_{\max}} + w_2 \varsigma_u(t) \\
    & + w_3 (BO_u(t) + EW_u(t)) \Big),
\end{split}
\end{equation}

Where coefficients $w_{u,1}, w_{u,2}, w_{u,3} \in [0, 1]$ represents the relative importance of buffer delay, carbon emissions, and energy wastage, respectively. The reward is defined as a negative weighted sum to penalize inefficient system behavior and guide the agent toward minimizing delay, energy inefficiency, and carbon emissions over time.

\subsection{Computational complexity and scalability analysis}
Computational complexity is evaluated in two phases: training and inference. The framework adopts parameter sharing, where actor and critic are fully connected DNNs with $L$ layers, $H$ neurons per hidden layer, state dimension $S$, and action dimension $A$.

During training, aggregated trajectories from all $U$ users update shared parameters using mini-batch SGD. For batch size $D$, backpropagation complexity is $O(D \cdot (S \cdot H + L \cdot H^2))$, scaling linearly with processed data. Although total training load grows with $U$, it is handled by edge/cloud servers, leaving on-device inference time unaffected during deployment. During the execution phase, each IoT device runs a local copy of the policy. The complexity of a forward pass is dominated by vector-matrix multiplications. For the first layer, the complexity is $O(S \cdot H)$, and for subsequent hidden layers, it is $O(H^2)$. Thus, the per-user inference complexity per time step is expressed as follows.
\begin{equation}
C_{\text{inference}} \approx O(S \cdot H + L \cdot H^2) \approx O(L \cdot H^2).
\end{equation}

Crucially, this complexity depends only on the local network size parameters $H$ and $L$ while remaining independent of the total number of users $U$. In a centralized DRL approach that concatenates the state space of all $U$ users, the input layer complexity would scale as $O(U \cdot S \cdot H + L \cdot H^2)$. In such a centralized model, the computational requirements grow linearly with the network size $U$. In contrast, the proposed decentralized CADDTO-PPO maintains a constant input size $S$ independent of $U$, yielding a per-device complexity of $O(L \cdot H^2)$ regardless of the network scale. This configuration confirms its suitability for future generations of massive IoT scenarios where adding devices does not increase the decision latency for existing users.
\section{Results and analysis}
\label{sec:results}

\subsection{Simulation setup}

The CADDTO-PPO framework is evaluated in a dynamic MEC setup using parameters from Table~\ref{tab:simulation_parameters}, with $U=5$ IoT devices by default. Users follow a random-walk mobility model within a $d = 100m$ BS radius. Channels are initialized as $v_u(0) \sim \mathcal{CN}(0, v_0(d_0/d_u)^\alpha I_Z)$ with correlation $\beta_u$ and error $e(t) \sim \mathcal{CN}(0, v_0(d_0/d_u)^\alpha I_Z)$. Mean task arrivals follow a Poisson process with $\lambda_u \in [2,10]$ bits, and energy harvesting is Poisson with $\lambda_E \in [1,5]$ Joules per slot. Harvested energy is prioritized, using grid power only when necessary. Carbon emissions are calculated using a grid intensity factor of $700$ g/CO$_2$ per kWh. State observations are normalized to $[0,1]$, and weightings are balanced $w_{u,1} = \frac{1}{3}, w_{u,2} = \frac{1}{3}, w_{u,3} = \frac{1}{3}$. Results are reproducible with seed 42. Experiments were run on an Intel Core i7-12700 CPU, 32 GB RAM, and NVIDIA RTX 3060 Ti GPU.

\begin{table}[t]
\footnotesize
\centering
\caption{System simulation and algorithm parameters}
\label{tab:simulation_parameters}
\begin{tabularx}{\columnwidth}{l X}
\toprule
\textbf{Parameter} & \textbf{Value} \\
\midrule
\multicolumn{2}{l}{\textbf{Network and Communication Settings}} \\
\midrule
Cell Radius ($R$) & $100$ m \\
Time Slot Duration ($\delta$) & $0.01$ s (10 ms) \\
System Bandwidth ($W$) & $1$ MHz \\
Noise Power ($\sigma^2$) & $1 \times 10^{-12}$ W \\
Path Loss Exponent ($\alpha$) & $2.0$ \\
Reference Distance ($d_0$) & $1$ m \\
Path Loss at $d_0$ & $-30$ dB \\
Temporal Correlation ($\rho$) & $0.95$ \\
\midrule
\multicolumn{2}{l}{\textbf{Hardware and Computing Settings}} \\
\midrule
Max. Transmit Power ($P_{o,u}$) & $2$ W \\
Max. Local Power ($P_{l,u}$) & $2$ W \\
CPU Cycles per bit ($N_u$) & $300$ cycles/bit \\
Effective Capacitance ($k$) & $1 \times 10^{-27}$ \\
Carbon Emission Factor ($e$) & $700$ g/kWh \\
MEC Processing Factor ($\eta$) & $1 \times 10^{-4}$ \\
\midrule
\multicolumn{2}{l}{\textbf{CADDTO-PPO Hyperparameters}} \\
\midrule
Learning Rate ($\alpha, \beta$) & $3 \times 10^{-4}$ \\
Discount Factor ($\gamma$) & $0.99$ \\
GAE Parameter ($\lambda$) & $0.95$ \\
PPO Clipping ($\epsilon$) & $0.2$ \\
Entropy Coefficient & $0.01$ \\
Mini-batch Size & $128$ \\
Buffer Size ($n_{steps}$) & $2048$ \\
\bottomrule
\end{tabularx}
\end{table}

\normalsize

\subsection{Performance evaluation}
To evaluate the effectiveness of the proposed CADDTO-PPO framework, we compare its performance against a diverse set of benchmark algorithms, including learning-based, optimization-based, and heuristic baselines. This is consistent with recent studies in MEC literature \cite{chenDecentralizedComputationOffloading2020,savazziEnergyCarbonFootprint2023,yangJointAccessSelection2025}.

\subsubsection{Benchmark algorithms}
\begin{itemize}
    \item \textbf{Centralized-PPO:} A fully centralized RL approach where a single agent at the BS observes the global system state of all users. This method serves as a theoretical upper bound on coordination capability but suffers from high computational complexity and limited scalability.
    
    \item \textbf{DRL-DDPG:} An independent Deep Deterministic Policy Gradient (DDPG) baseline representing off-policy deterministic control \cite{chenDecentralizedComputationOffloading2020,weiDeepReinforcementLearning2023}. This benchmark evaluates the effectiveness of deterministic action execution against the proposed on-policy formulation.
    
    \item \textbf{Lyapunov-DPP:} A classical optimization-based control policy that minimizes the drift-plus-penalty function at each time slot. While effective for queue stability, it lacks long-term learning capability and relies solely on instantaneous state information.
    
    \item \textbf{Heuristic Baselines:} These include \textit{Greedy} (maximum power allocation), \textit{Local-only} (no task offloading), and \textit{Offload-only} (no local execution), which represent non-learning heuristic baselines that prioritize immediate throughput or energy efficiency without considering long-term system dynamics or trade-offs.
\end{itemize}

\begin{figure}[t]
    \centering
    \includegraphics[width=\columnwidth]{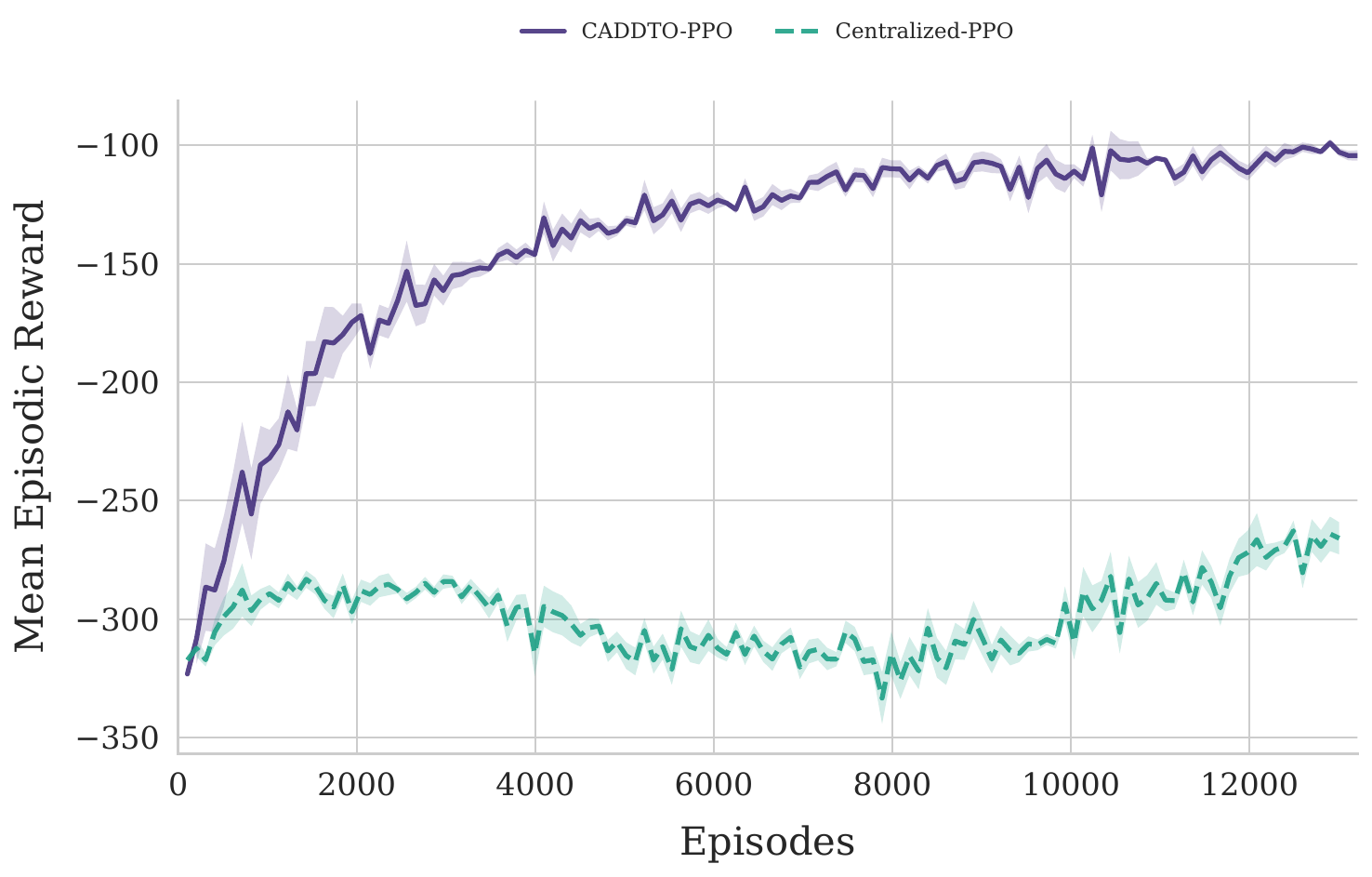}
    \caption{Training convergence of CADDTO-PPO and Centralized-PPO.}
    \label{fig:ppo_training}
\end{figure}

\subsection{Convergence of CADDTO-PPO}

The training and convergence of CADDTO-PPO are compared with Centralized-PPO as shown in Fig. \ref{fig:ppo_training} over $1.3 \times 10^6$ steps (13,000 episodes, $T_{\max}=100$) with task arrival $\lambda_u=4$ bits. CADDTO-PPO steadily improves from an initial reward of $-325$ to a stable plateau near $-100$, adapting effectively using only local observations. Its stability stems from decentralized execution with parameter sharing, which mitigates non-stationarity, while narrow variance bands indicate reliable gradient estimation via PPO clipped updates. In contrast, Centralized-PPO stagnates around $-300$. Managing the joint state-action space of all $U$ users leads to the Curse of Dimensionality, high-variance gradients, and performance degradation, compounded by the non-stationary hybrid-energy environment, preventing convergence despite global CSI access.

\begin{figure*}[t]
    \centering
    \begin{subfigure}[b]{0.32\textwidth}
        \centering
        \includegraphics[width=\textwidth]{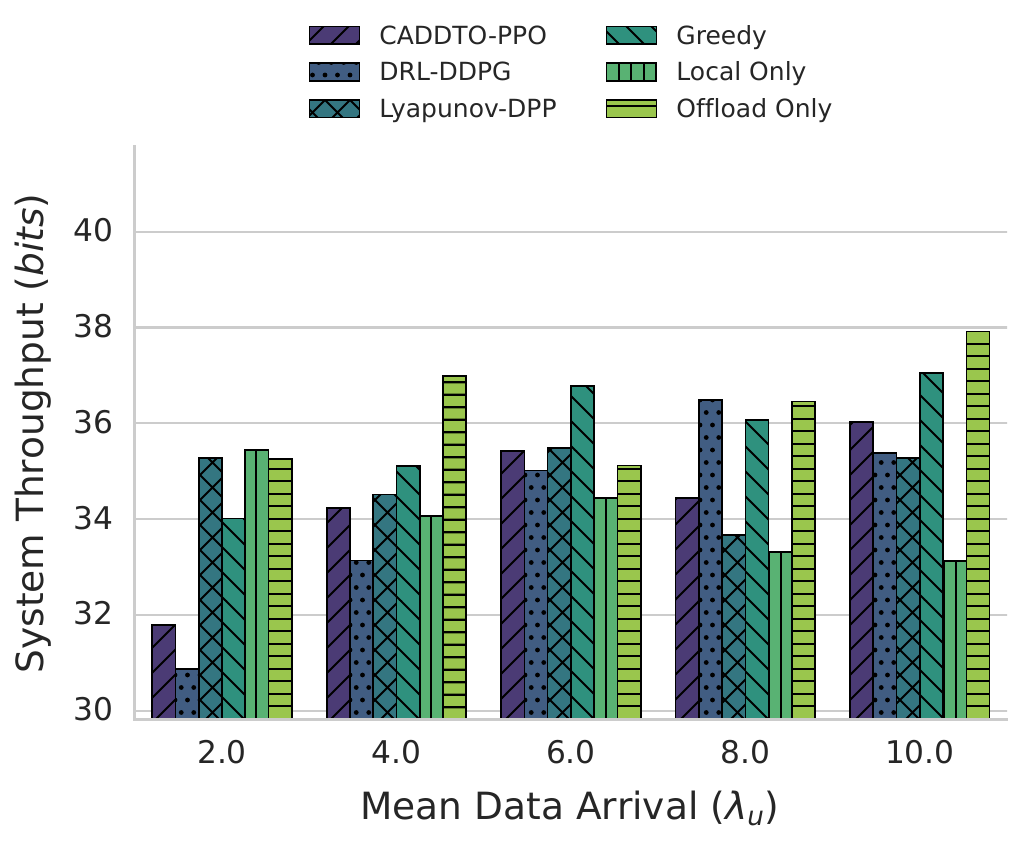}
        \caption{System Throughput}
        \label{fig:throughput}
    \end{subfigure}
    \hfill
    \begin{subfigure}[b]{0.32\textwidth}
        \centering
        \includegraphics[width=\textwidth]{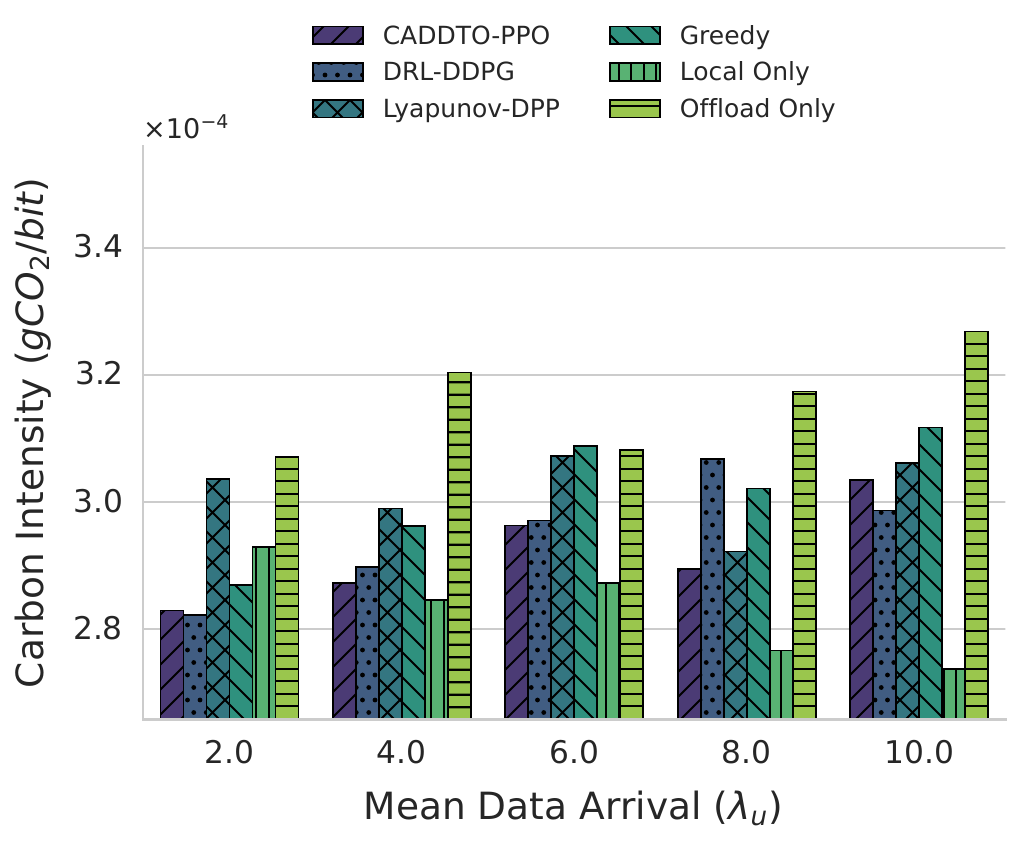}
        \caption{Carbon Intensity}
        \label{fig:carbon}
    \end{subfigure}
    \hfill
    \begin{subfigure}[b]{0.32\textwidth}
        \centering
        \includegraphics[width=\textwidth]{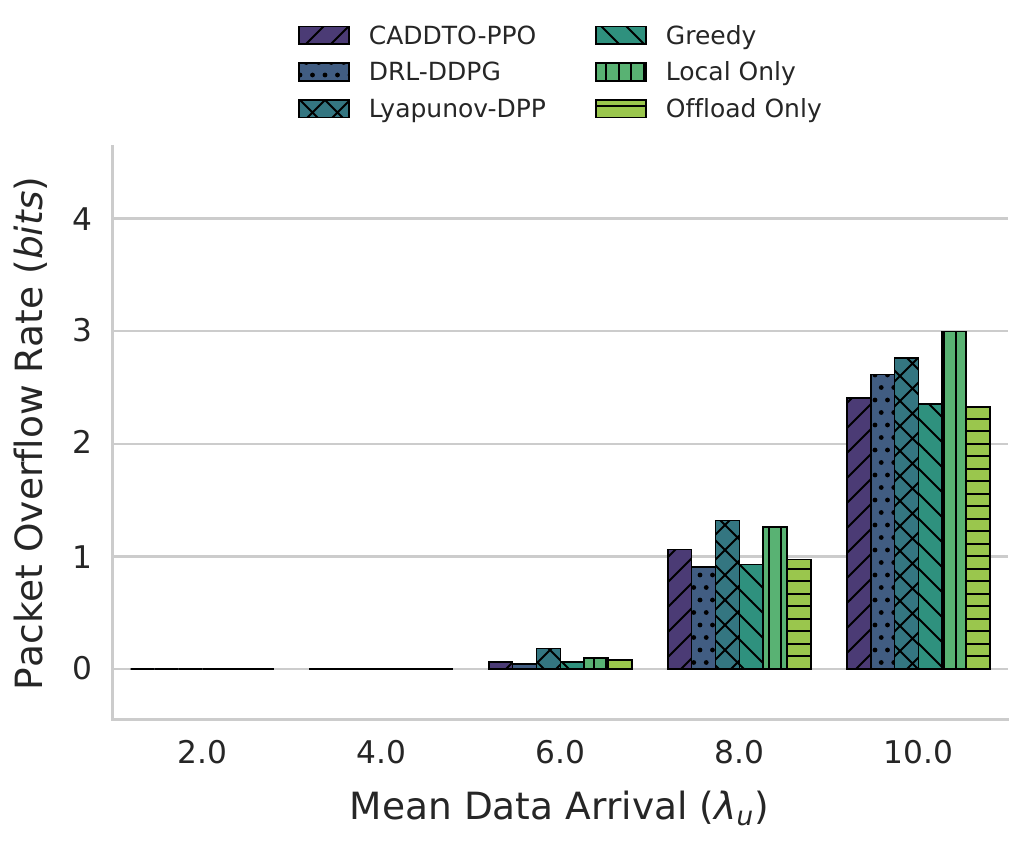}
        \caption{Packet Overflow Rate}
        \label{fig:overflow}
    \end{subfigure}
    \caption{Performance evaluation under varying mean data arrival rates ($\lambda_{u} \in [2, 10]$ bits).}
    \label{fig:arrival_analysis}
\end{figure*}

\subsection{Impact of traffic load on eco-Efficiency and stability}
In this section, we evaluate the impact of varying the mean data arrival rate, $\lambda_{\text{u}}$, on three critical performance metrics: system throughput, carbon intensity, and packet overflow rate. To ensure statistical significance, all results are averaged over 100 independent simulation runs.

As illustrated in Fig.~\ref{fig:throughput}, system throughput rises with mean arrival rates from 2 to 10 bits, indicating effective MEC resource utilization. While Offload Only and Greedy policies achieve the highest raw throughput at ($\lambda_{\text{u}}=10bits$) bits, this comes with high carbon emissions. CADDTO-PPO maintains competitive throughput, within 5\% of Greedy, outperforming Lyapunov-DPP, whose myopic control limits adaptation under fluctuating multi-user interference. DRL-DDPG achieves similar throughput but with higher variance. CADDTO-PPO eco-efficiency policy occasionally throttles processing to prevent battery depletion or grid-energy penalties during poor channel conditions.

The environmental impact is analyzed through the carbon intensity metric (gCO$_2$/bits), shown in Fig.~\ref{fig:carbon}. Unlike heuristic policies and other RL baselines, CADDTO-PPO achieves the lowest gCO$_2$/bit across all arrival rates. Offload Only spikes above 0.00032~gCO$_2$/bit due to constant grid use, while DRL-DDPG fails to adapt to interference fluctuations, increasing grid reliance. At $\lambda_u=10$ bits, CADDTO-PPO reduces carbon per bit by ~15.6\% versus Offload Only, confirming effective prioritization of green slots. Fig.~\ref{fig:overflow} depicts the packet overflow rate, which serves as a proxy for system reliability and queue stability. For $\lambda_u < 6$ bits, all policies maintain near-zero overflow. Beyond $\lambda_u \ge 8$ bits, overflow rises sharply. The Lyapunov-DPP policy and Local Only scheme exhibit the highest overflow rates at $\lambda_{\text{u}}=10bits$ (approximately 2.7 and 3.0 bits, respectively). DRL-DDPG manages to keep overflow lower than heuristic baselines, it still results in higher packet loss compared to CADDTO-PPO under extreme congestion. By including buffer state in observations and penalizing overflow, CADDTO-PPO proactively offloads tasks, achieving the lowest overflow among learning-based policies.

\begin{figure*}[t]
    \centering
    \begin{subfigure}[b]{0.32\textwidth}
        \centering
        \includegraphics[width=\textwidth]{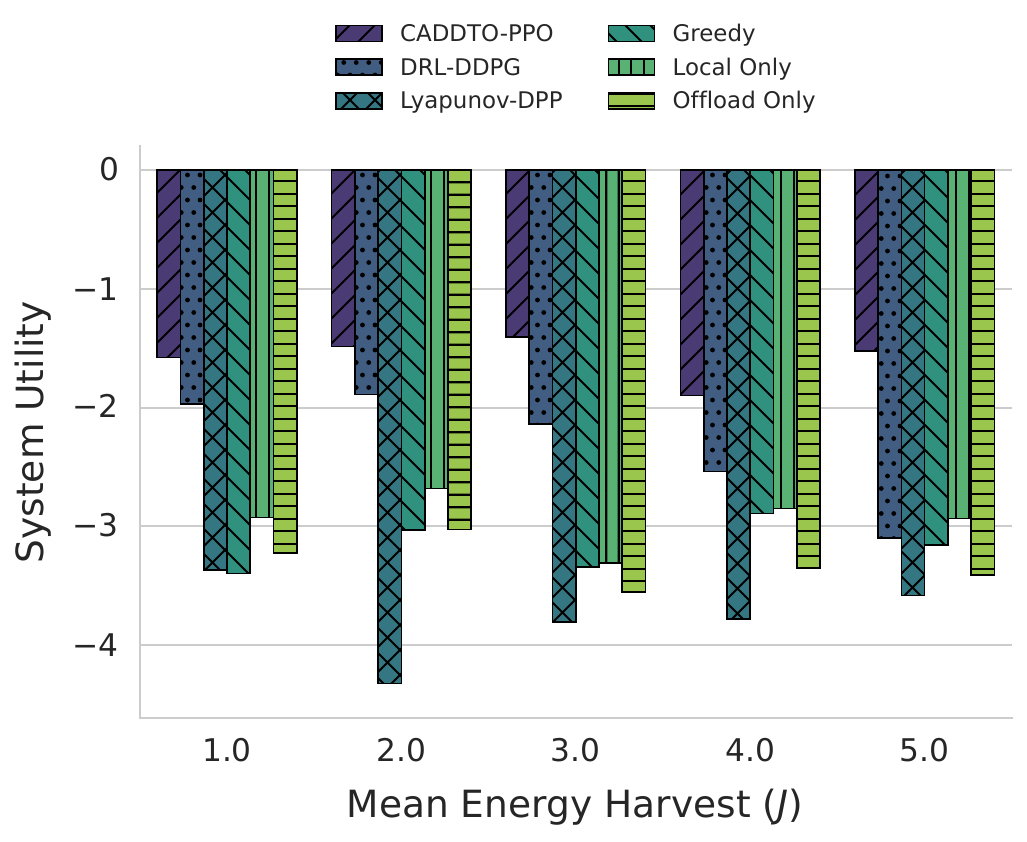}
        \caption{System Utility}
        \label{fig:energy_reward}
    \end{subfigure}
    \hfill
    \begin{subfigure}[b]{0.32\textwidth}
        \centering
        \includegraphics[width=\textwidth]{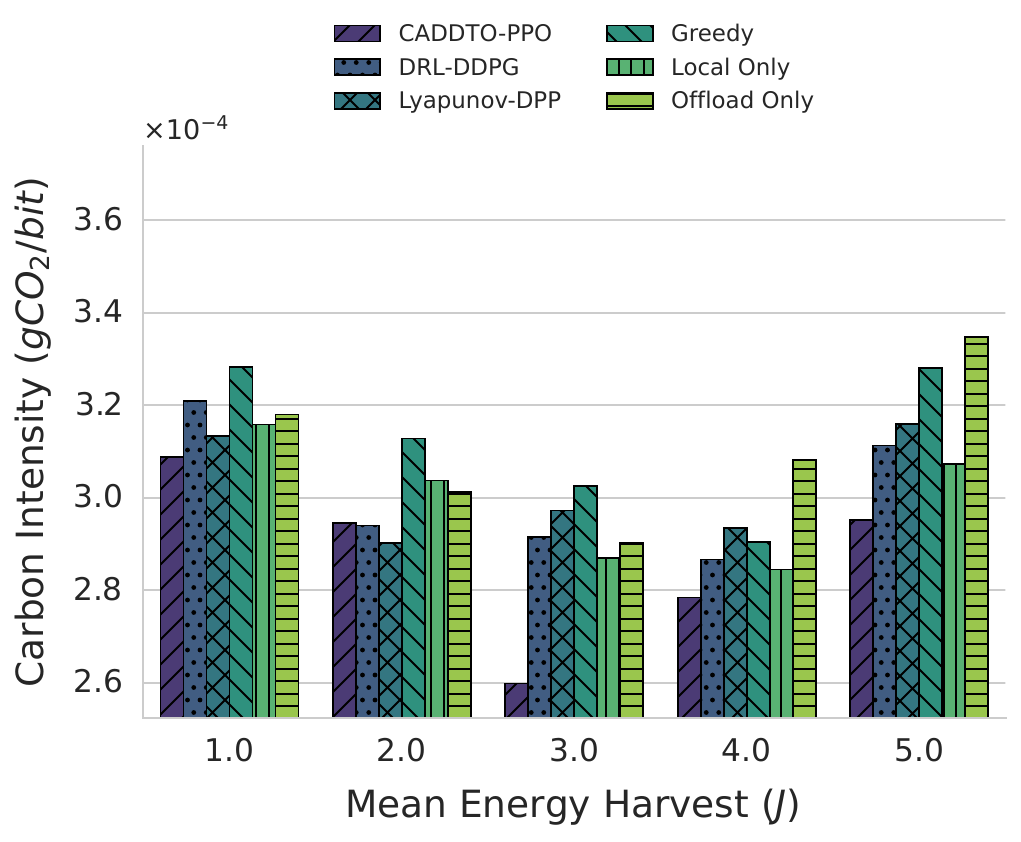}
        \caption{Carbon Intensity}
        \label{fig:energy_carbon}
    \end{subfigure}
    \hfill
    \begin{subfigure}[b]{0.32\textwidth}
        \centering
        \includegraphics[width=\textwidth]{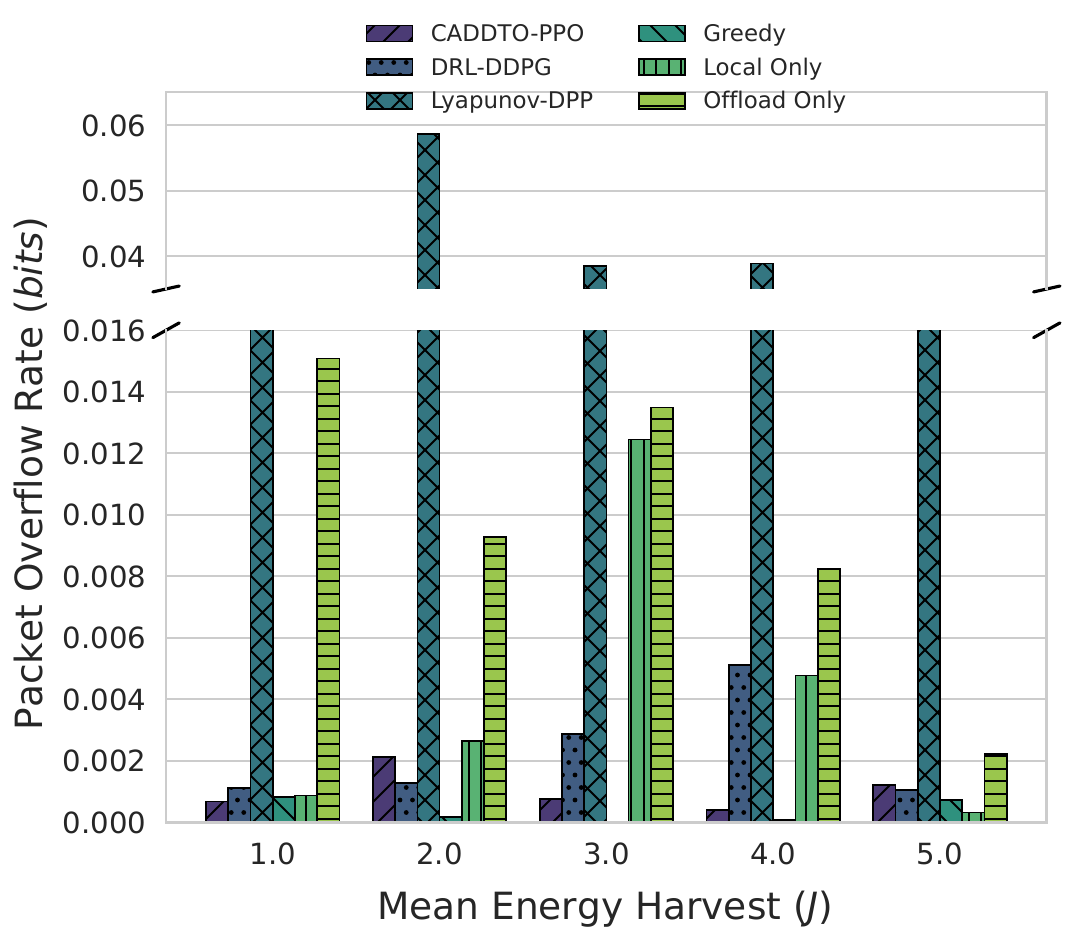}
        \caption{Packet Overflow Rate}
        \label{fig:energy_overflow}
    \end{subfigure}
    \caption{Sustainability analysis under varying mean energy harvest rates ($\lambda_E \in [1, 5]$ Joules).}
    \label{fig:sustainability_analysis}
\end{figure*}

\subsection{Impact of renewable energy availability}
This section analyzes the impact of renewable energy availability on system performance. By varying the mean energy harvest rate $\lambda_{E}$ from 1.0 to 5.0 Joules, we evaluate how the decentralized agents adapt their power allocation strategies to utilize stochastic green energy. The results are presented in Fig.~\ref{fig:sustainability_analysis} across three metrics: system utility, carbon intensity, and packet overflow.

The average system utility, which represents the aggregate multi-objective reward, is illustrated in Fig.~\ref{fig:energy_reward}. CADDTO-PPO consistently achieves the highest utility$-1.5$ and $-2.0$ across all energy harvesting levels, demonstrating strong adaptability to stochastic energy arrivals. DRL-DDPG ranks second but degrades under unpredictable harvesting due to stale transitions in its replay buffer, whereas CADDTO-PPO benefits from on-policy updates using recent trajectories.
Lyapunov-DPP exhibits lower and more volatile utility, as it relies on instantaneous measurements rather than long-term energy patterns. Heuristic baselines underperform overall: Greedy yields the lowest utility due to high energy and carbon penalties, while Local Only and Offload Only remain stagnant, failing to exploit increased harvested energy.

Fig.~\ref{fig:energy_carbon} shows carbon intensity (gCO$_2$/bit) as a measure of environmental efficiency. As mean harvested energy increases, CADDTO-PPO exhibits a clear downward trend, indicating a shift from grid dependence to self-sustained operation. DRL-DDPG improves with higher energy availability but maintains higher carbon intensity due to Q-value overestimation and aggressive power use. Lyapunov-DPP and Greedy policies incur inefficient power usage, relying on instantaneous drift control or persistent grid consumption. At $\lambda_{E}=3.0$, CADDTO-PPO achieves the lowest carbon intensity, outperforming DRL-DDPG and Offload Only. By dynamically balancing local execution and offloading to prioritize green slots, CADDTO-PPO more effectively decouples throughput from grid energy consumption than deterministic baselines.

The reliability of the system under varying energy budgets is evaluated through the packet overflow rate in Fig.~\ref{fig:energy_overflow}. Due to the extreme performance disparity, a broken-axis plot is utilized to highlight the stability of learning-based methods against the baseline outliers. Lyapunov-DPP is identified as a massive outlier, indicating frequent buffer saturation. While DRL-DDPG maintains relatively stable queues, it still exhibits higher overflow than CADDTO-PPO, particularly when energy harvesting is low ($\lambda_{E}=1.0$). This failure in DDPG is due to its inability to strictly enforce the stability constraints when the policy gradient becomes noisy under low-energy conditions. Conversely, CADDTO-PPO maintains an overflow rate near zero for all energy levels. By penalizing energy wastage and buffer overflow, the agent learns a proactive drainage policy, allocating higher transmission power when the buffer is nearing capacity even if harvested energy is low.

\begin{figure*}[t]
    \centering
    \begin{subfigure}[b]{0.32\textwidth}
        \centering
        \includegraphics[width=\textwidth]{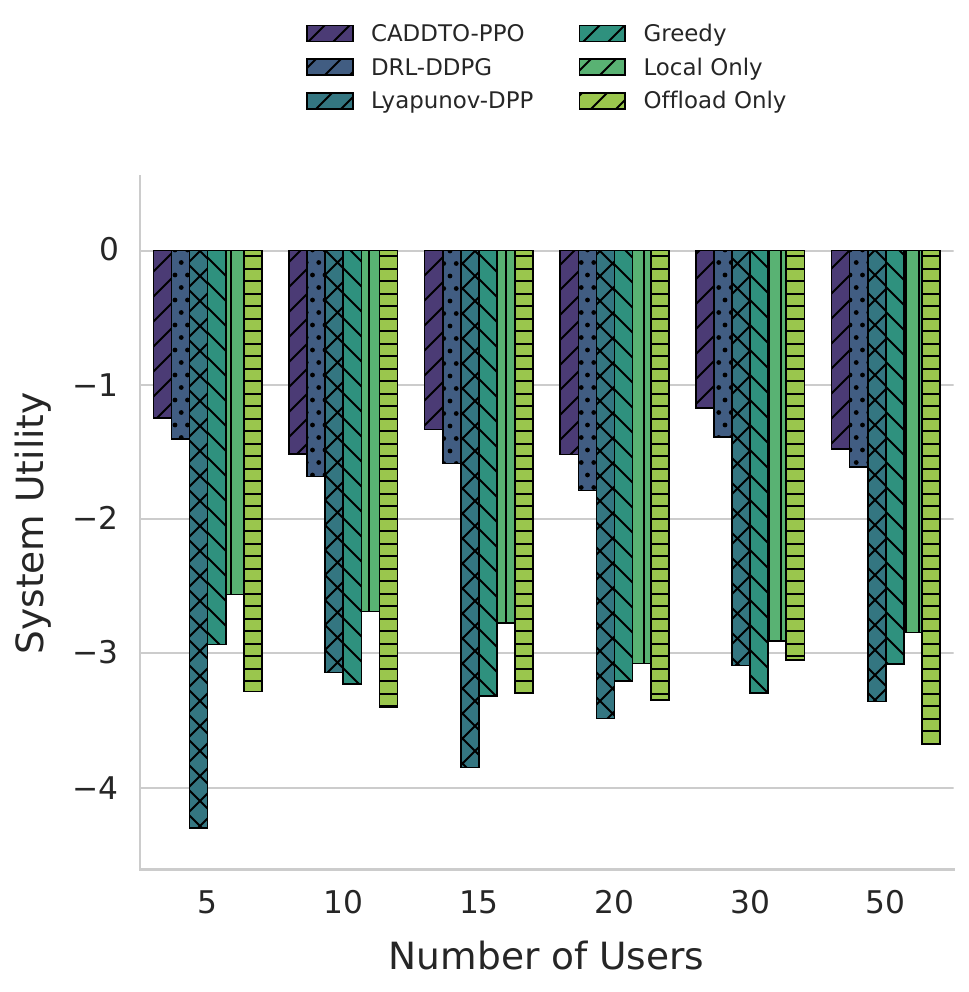}
        \caption{System Utility}
        \label{fig:scale_reward}
    \end{subfigure}
    \hfill
    \begin{subfigure}[b]{0.32\textwidth}
        \centering
        \includegraphics[width=\textwidth]{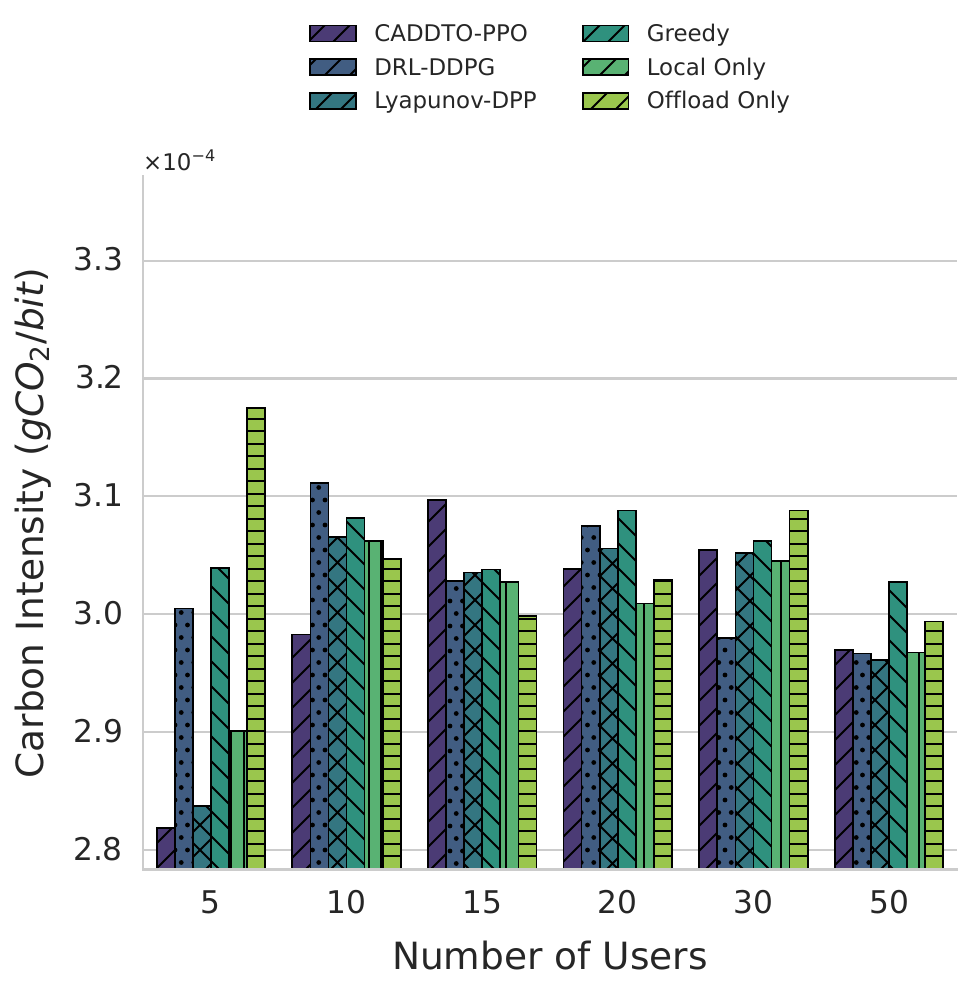}
        \caption{Carbon Intensity}
        \label{fig:scale_carbon}
    \end{subfigure}
    \hfill
    \begin{subfigure}[b]{0.32\textwidth}
        \centering
        \includegraphics[width=\textwidth]{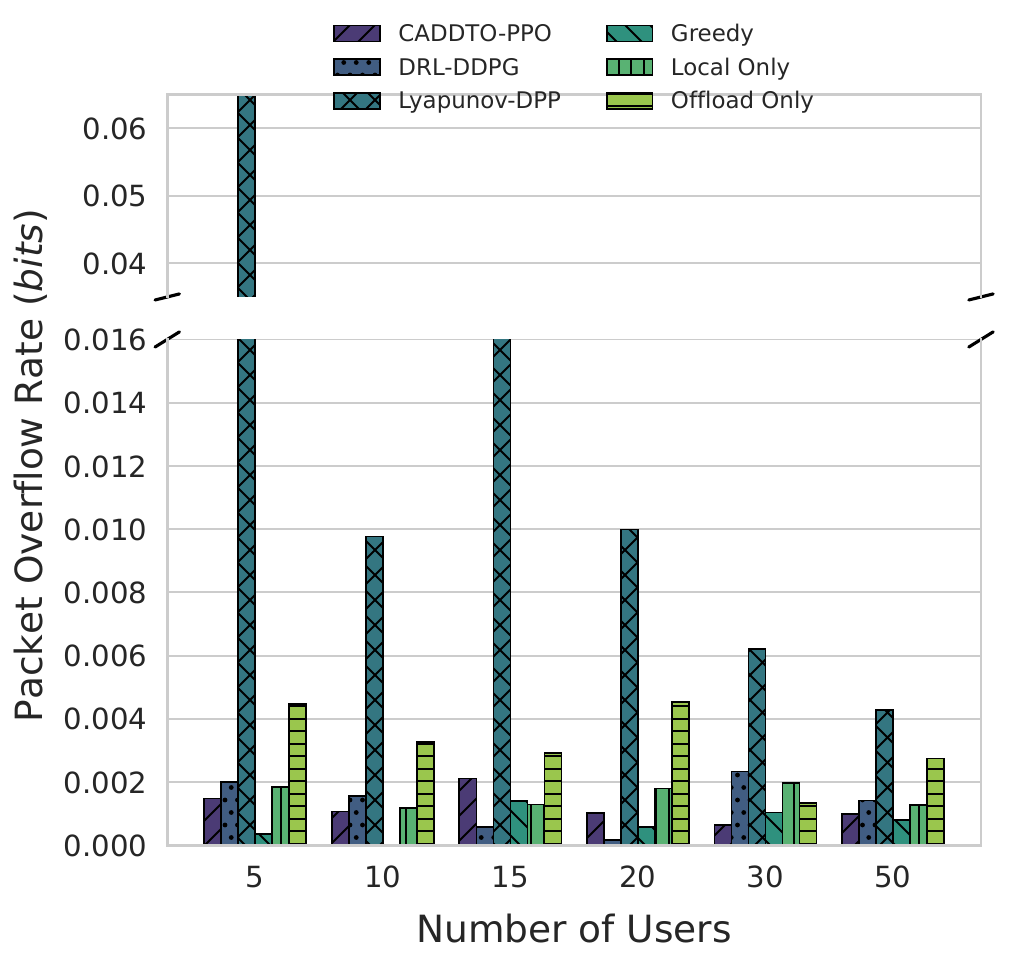}
        \caption{Packet Overflow Rate}
        \label{fig:scale_overflow}
    \end{subfigure}
    \caption{Scalability analysis across varying user densities ($U \in [5, 50]$).}
    \label{fig:scalability_analysis}
\end{figure*}

\subsection{Scalability and interference robustness}


This section evaluates the scalability and robustness of the proposed framework as IoT device density increases from 5 to 50 users. A critical factor for massive IoT networks facing co-channel interference Fig.~\ref{fig:scalability_analysis}. The average system utility across varying user counts is shown in Fig.~\ref{fig:scale_reward} shows that the CADDTO-PPO policy remains stable, maintaining high system utility ($\approx -1.2$ to $-1.5$) as the network scales. Unlike centralized and off-policy RL methods, which degrade with more agents, and Lyapunov-DPP, which drops significantly at $U=15$, proposed decentralized MAPPO with parameter sharing effectively manages multi-user complexity. The flat utility trend demonstrates that decentralized agents sustain optimal performance, highlighting the framework suitability for dense urban and industrial IoT deployments.

Fig.~\ref{fig:scale_carbon} analyzes the carbon intensity per bits as the number of users grows. Ideally, a scalable system should maintain a constant or decreasing carbon intensity even as the aggregate workload increases. CADDTO-PPO successfully achieves this, with the carbon intensity remaining stable near 0.00030~gCO$_2$/bits across all scales. This eco-scalability is a direct result of the decentralized decision-making process; since each agent optimizes its power allocation based on local observations and experienced interference, the system avoids the race to the top power wars typical in greedy or uncoordinated networks. Heuristic policies like Offload Only show an upward trend in carbon intensity as user density increases, as the rising interference necessitates higher transmission power to maintain data rates, whereas CADDTO-PPO manages these trade-offs through intelligent local-offloading switching.

System reliability under high interference is evaluated via packet overflow Fig.~\ref{fig:scale_overflow}. As user count rises, MIMO uplinks become crowded, yet CADDTO-PPO maintains overflow below 0.002 bits even at $U=50$, far outperforming baselines. Lyapunov-DPP spikes to 0.07 bits at $U=5$ and remains erratic at higher densities. CADDTO-PPO robustness derives from $O(1)$ per-device inference and its interference-aware policy, which proactively drains buffers without global CSI, preventing network-wide congestion as the system scales.

\begin{figure}[t]
    \centering
    \includegraphics[width=\columnwidth]{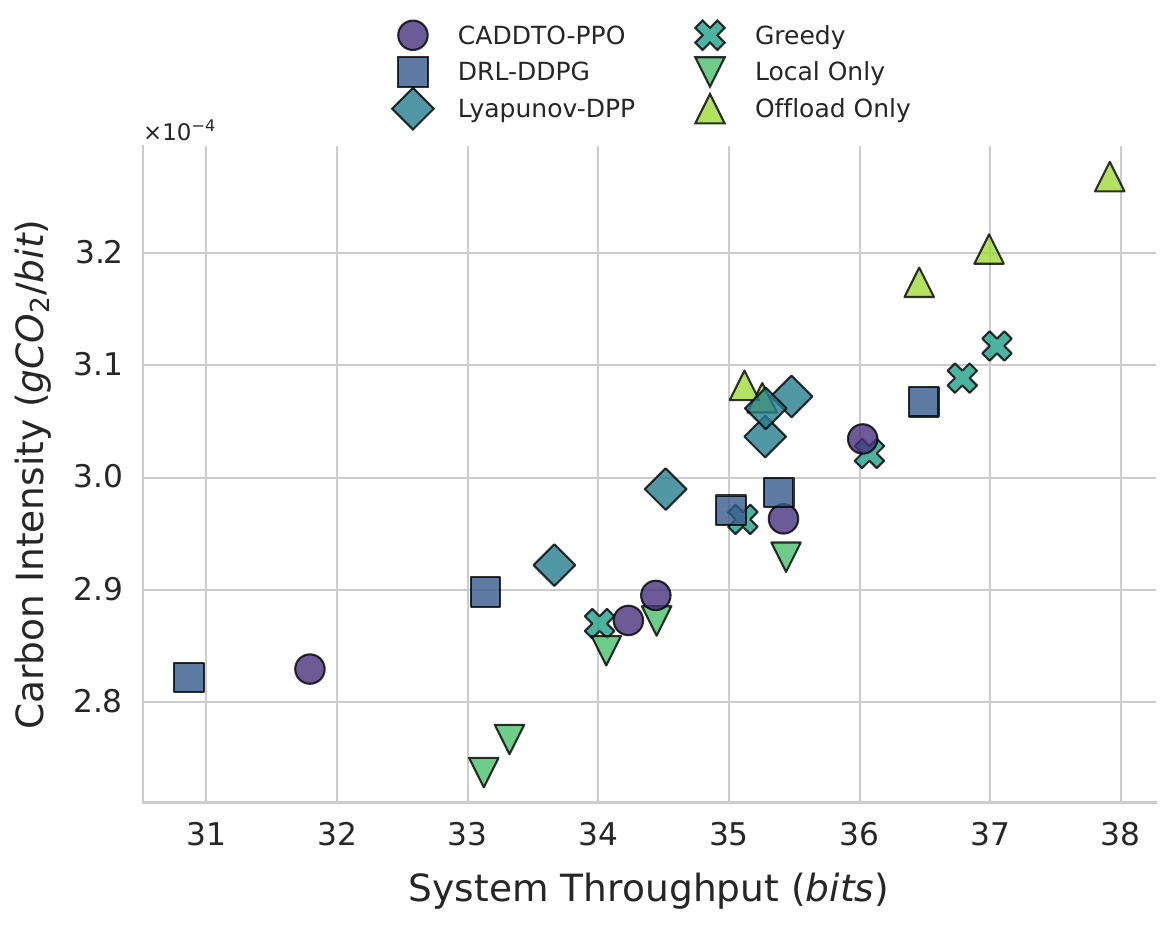}
    \caption{Eco-efficiency trade-off analysis: System throughput vs. carbon intensity.}
    \label{fig:tradeoff_analysis}
\end{figure}

\subsection{Multi-objective eco-efficiency trade-off analysis}

The multi-objective performance is evaluated through the eco-efficiency trade-off between throughput and carbon intensity Fig.~\ref{fig:tradeoff_analysis}. This Pareto-style visualization highlights how policies balance data processing and environmental impact under varying traffic and energy conditions.Each marker represents the average performance of a specific policy under varying traffic and energy conditions. CADDTO-PPO occupies the favorable lower-right region, achieving throughput above 35 bits while keeping carbon intensity below 0.00030 gCO$_2$/bits. This reflects its ability to learn long-term interactions among interference, renewable energy, and grid penalties. In contrast, Offload Only attains the highest throughput 38 bits but with the worst carbon efficiency, exceeding 0.00032 gCO$_2$/bit, highlighting its unsustainable reliance on grid-powered transmission.

DRL-DDPG exhibits unstable performance across the trade-off space, with scattered points ranging from low-throughput/low-carbon to high-throughput/high-carbon regions. This dispersion reflects its difficulty in maintaining a consistent balance in high-interference MIMO settings. Although DDPG occasionally matches CADDTO-PPO carbon levels, it often sacrifices throughput, dropping near 31 bits. Lyapunov-DPP clusters at moderate performance due to its myopic allocation strategy, while Local Only remains in the low-throughput/low-carbon region, constrained by device CPU limits. The Greedy policy attains high throughput but incurs higher carbon cost per bit. Overall, CADDTO-PPO multi-objective reward enables operation along the optimal eco-efficiency frontier, delivering a more sustainable solution for massive IoT networks.

\subsection{Architectural profiling and edge feasibility analysis}
To validate the practical applicability of CADDTO-PPO in resource-constrained MEC environments, we conducted a comprehensive hardware profiling analysis. The architectural complexity was evaluated using standard multiply-accumulate (MAC) and floating point operations (FLOPs) metrics where $1 \text{ MAC} \approx 2 \text{ FLOPs}$. As summarized in Table~\ref{tab:hardware_profile}, the proposed agent requires only $17.02 \text{ K}$ MACs and $34.05 \text{ K}$ FLOPs per inference. Serializing the model into the ONNX format yields an exceptionally compact footprint of $8.29 \text{ KB}$ with a runtime static memory requirement of $67.51 \text{ KB}$ for float32 parameters. This ultra-lightweight profile enables seamless deployment across a spectrum of hardware ranging from low-power ARM microcontrollers and ESP32 SoCs to Linux-based edge gateways like the Raspberry Pi where the model can reside entirely within the L1 data cache to minimize DRAM access latency.

Performance was benchmarked on an Intel Xeon CPU (2.20GHz) in single-threaded mode to provide a conservative estimate for resource-constrained IoT processors. CADDTO-PPO achieved an average inference latency of $0.1457 \text{ ms}$, representing a $1.36\times$ speedup over the Lyapunov-DPP optimizer ($0.1981 \text{ ms}$). Unlike the iterative Lyapunov heuristic, which suffers from exponential complexity $O(G^A)$ relative to the search granularity $G$ and action dimensions $A$, CADDTO-PPO provides constant-time inference relative to the action search space. While its internal operations follow the architecture-dependent complexity of $O(L \cdot H^2)$, the execution time remains invariant to the precision or range of the action space. Given a system time-slot of $\tau = 10 \text{ ms}$, the proposed agent utilizes only $1.45\%$ of the temporal budget. This high computational efficiency is particularly beneficial for energy-harvesting nodes, as rapid decision-making allows the processor to transition to a low-power sleep state sooner, thereby significantly extending the operational longevity of self-sustained MEC networks.

\begin{table}[t]
\footnotesize
\centering
\caption{Comparative hardware profiling and deployment metrics}
\label{tab:hardware_profile}
\begin{tabularx}{\columnwidth}{l X X}
\toprule
\textbf{Metric} & \textbf{CADDTO-PPO} & \textbf{Lyapunov-DPP} \\
\midrule
Network architecture & MLP [128, 128] & Logic Search \\
Total parameters & 17.28 K & N/A \\
MAC operations & 17.02 K & N/A \\
FLOP operations & 34.05 K & N/A \\
Storage size (ONNX) & 8.29 KB & N/A \\
Runtime RAM usage & 67.51 KB & Low \\
Inference latency & 0.1457 ms & 0.1981 ms \\
Time-slot utilization & 1.45\% & 1.98\% \\
\bottomrule
\end{tabularx}
\end{table}

\section{Conclusion and future work}

This work addresses decentralized, carbon-aware task offloading in multi-user MIMO-enabled MEC systems powered by hybrid renewable energy. We propose CADDTO-PPO, a scalable Multi-Agent-based framework with parameter sharing that enables IoT agents to learn power control and offloading policies from local observations in interference and energy-uncertain environments. Results show that CADDTO-PPO effectively captures the interference-carbon coupling, outperforming Lyapunov-DPP and DRL-DDPG baselines. It achieves the lowest carbon intensity ($gCO_2/bit$) with near-zero packet overflow under heavy traffic ($\lambda_u = 10$ bits/slot). Scalability analysis confirms constant-time decision complexity from 5 to 50 users, while hardware profiling demonstrates negligible inference latency (<1 ms) and minimal memory usage, validating real-world feasibility.

However, our current modeling primarily emphasizes a single-cell environment where IoT devices remain within the coverage of a dedicated base station. Future work will extend the model to multi-cell scenarios with mobility and handovers, incorporate dynamic grid carbon intensity, and explore integration with semantic communication (SemCom) to reduce uplink load. We also plan to design decentralized incentive mechanisms to sustain long-term participation in collaborative green edge computing.

\section*{CRediT authorship contribution statement}
\textbf{Mubshra Zulfiqar:} Writing – review \& editing, Writing – original draft, Methodology, Investigation, Formal analysis, Data curation, Conceptualization; \textbf{Muhammad Ayzed Mirza:} Validation, Project administration, Methodology, Review; \textbf{Basit Qureshi:}  Supervision, Resources, Project administration, Methodology, Review, Funding acquisition.

\section*{Data availability}
Data will be made available on request.

\section*{Declaration of competing interest}
The authors declare that they have no known competing financial interests or personal relationships that could have appeared to influence the work reported in this paper.

\section*{Acknowledgments}
This research work is partially supported by Prince Sultan University, Saudi Arabia.

 \bibliographystyle{elsarticle-num} 
 \bibliography{biblography}


\begin{wrapfigure}{l}{0.22\textwidth}
    \vspace{-15pt}
    \includegraphics[width=0.22\textwidth]{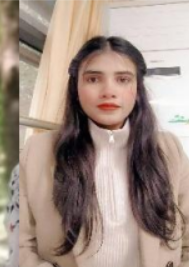}
    \vspace{-20pt}
\end{wrapfigure}
\textbf{Mubshra Zulfiqar} earned her Masters from Wuhan Textile University in 2021 and is pursuing a Ph.D. at the Wuhan University of Technology. Her research focuses on MEC, Deep Learning, Power-aware Computing, task offloading and the IoT.

\vspace{1.5em} 

\begin{wrapfigure}{l}{0.22\textwidth}
    \vspace{-15pt}
    \includegraphics[width=0.22\textwidth]{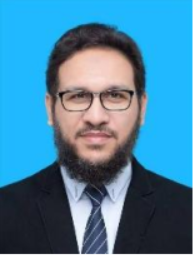}
    \vspace{-20pt}
\end{wrapfigure}
\textbf{Muhammad Ayzed Mirza} earned his MSCS from National Textile University in 2016. He completed his Ph.D. in 2023 at Beijing University of Posts and Telecommunications. He is now an Associate Professor at Qilu Institute of Technology. His research focuses on B5G/6G Vehicular Edge Computing Networks, task offloading, RIS/NOMA communications.

\vspace{1.5em}

\begin{wrapfigure}{l}{0.22\textwidth}
    \vspace{-15pt}
    \includegraphics[width=0.22\textwidth]{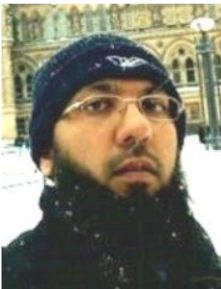}
    \vspace{-20pt}
\end{wrapfigure}
\textbf{(Senior Member, IEEE)} is a Full Professor in the Department of Computer Science at the College of Computer and Information Sciences, Prince Sultan University, Saudi Arabia. He received his BSc Degree in Computer Science from Ohio University, OH USA in 2000 followed by a MSc degree in Computer Science from Florida Atlantic University, FL USA in 2002 and Ph.D. degree from the University of Bradford UK in 2011. His research focuses on cloud computing, with particular emphasis on energy-aware and green computing, performance optimization, and the integration of cloud and IoT technologies for smart city applications. He has authored over 100 peer-reviewed publications in leading international journals and conferences and received the Best Paper Award at the IEEE/IFIP TrustCom 2010 Conference in Hong Kong. Dr. Qureshi has served on the organizing and program committees of numerous international conferences and has been the Chair and Co-Chair of the Series of International Conference on Data Science and Machine Learning (CDMA), held biennially in Saudi Arabia since 2010. He previously served as Director of the Prince Megrin Data Mining Center and Chair of the Computer Science Department at Prince Sultan University. He is a Fellow of the UK Higher Education Academy UK and a Senior Member of IEEE, IEEE CS and ACM.

\end{document}